\documentclass[aps,prb,twocolumn,secnumarabic,amsmath,amssymb,superscriptaddress]{revtex4-1}
\usepackage{dcolumn}
\usepackage{bm}

\usepackage{color} 

\usepackage{ulem}

\usepackage{amsmath}
\usepackage{graphicx}
\usepackage{float}
\usepackage{subfig}

\usepackage{color}
\usepackage[colorlinks,bookmarks=false,citecolor=blue,linkcolor=red,urlcolor=blue]{hyperref}

\definecolor{darkred}{rgb}{0.7,0.0,0.0}

\definecolor{darkblue}{rgb}{0,0.02,0.45}

\definecolor{darkgreen}{rgb}{0.02,0.45,0.0}

\definecolor{violet}{rgb}{0.8,0.2,0.6}

\providecommand{\U}[1]{\protect\rule{.1in}{.1in}}

\begin{document}

\title{Majorana representations of spin and\\ an alternative solution of the Kitaev honeycomb model}

\author{Jianlong Fu}
\affiliation{School of Physics and Astronomy, University of Minnesota, Minneapolis,
Minnesota 55455, USA}
\author{Johannes Knolle}
\affiliation{Blackett Laboratory, Imperial College London, London SW7 2AZ, United Kingdom}
\author{Natalia  B. Perkins}
\affiliation{School of Physics and Astronomy, University of Minnesota, Minneapolis,
Minnesota 55455, USA}

\begin{abstract}
Based on the Dirac spinor representation of the SO(4) group, we discuss the relationship between three types of representation of spin in terms of Majorana fermions, namely the Kitaev representation, the SO(3) representation and the SO(4) chiral representation. Comparing the three types, we show that the Hilbert space of the SO(3) representation is different from the other two by requiring pairing of sites, but it has the advantage over the other two in that no unphysical states are involved. As an example of its application, we present a new alternative solution of the Kitaev honeycomb model. Our new solution involves no unphysical states which enables a systematic calculation of physical observables. Finally, we discuss an extension of the model to a more general exactly soluble $Z_{2}$ gauge theory interacting with complex fermions. 
\end{abstract}

\maketitle
\section{Introduction}\label{introduction}

The study of quantum spin liquid (QSL) states, which are exotic ground states of spin systems, has become one of the central problems of interest in the field of strongly correlated electrons.\cite{Anderson1973,FazekasAnderson74,Kalmeyer1987,Wen1989,Moessner2001, Kitaev2003,Kitaev2006,Balents2010,Savary2016,Zhou2017} The interest in QSLs originates from their novel properties including the absence of magnetic long-range order even at zero temperature, long-range quantum entanglement, topological ground-state degeneracy, and fractionalized excitations. Most of these exotic behaviors cannot be captured by traditional perturbative approaches, instead, in many cases the understanding of QSL ground states and the corresponding low-energy excitations has to rely on numerical or variational techniques.\cite{LiangDoucotAnderson88,YanHuseWhite2011,Shollwock2012,Poilblanc2012,Balents2012,Wan2013,IoannisZ2}    
On the other hand, substantial theoretical understanding of QSLs has been obtained by slave-particle approaches, in which the fractionalization of elementary spin flip excitations is taken into account by construction. \cite{Wen1989,Arovas88,Wen1991,Lee2006,Plee2011,Savary2016}  In particular the slave-particle approaches involve the representation of spins in terms of fractionalized bosonic or fermionic degrees of freedom (dubbed partons). The most commonly used is the complex fermion (also called Abrikosov fermion) representation. \cite{Savary2016,Lee2006} However, as the resulting Hamiltonian is usually quartic in partons, in most cases the analysis of the ground state and the excitation spectrum can be performed only at the mean-field level. Although this approach can be justified in some limits, its applicability to physical spin models is questionable in many cases. 

In rare cases, exact solutions of QSL Hamiltonians are achievable. The solutions not only enable the rigorous study of the QSL properties themselves,~\cite{Knolle2014,Knolle2014b,Knolle2015} but also provide us with opportunities to compare the applicability of other approximate theoretical methods.~\cite{Burnell2011,Knolle2018} Of particular interest in this regard is the Kitaev honeycomb model which describes a system of spin-1/2 at sites of a honeycomb lattice interacting via Ising-like nearest-neighbor exchange interactions.\cite{Kitaev2006} This model is not only exactly solvable with a QSL ground state, but also realizable in materials.\cite{Kitaev2006,Baskaran2007,Jackeli2009} In particular, recent years have seen much progress in identifying candidate materials
for realizing the Kitaev QSL, such as the honeycomb iridates  A$_2$IrO$_3$ (with A$=$Na/Li), \cite{Singh2010,Singh2012,Ye2012,Chun2015,Williams2016} honeycomb ruthenium chloride $\alpha-$RuCl$_3$, \cite{Plumb2014,Sears2015,Johnson2015} and in another 5d Ir honeycomb compound H$_3$LiIr$_2$O$_6$.\cite{Takagi2018} We refer interested readers to some recent reviews on the development of Kitaev materials.\cite{Krempa2014,Rau2016,Trebst2017,Knolle2017} 

The exact solution of the Kitaev honeycomb model has been achieved by a slave-particle representation of spins using four Majorana fermions in an extended Hilbert space.\cite{Kitaev2006} Using this representation, Kitaev demonstrated that the low-energy physics of the original spin model can be understood by studying a system of Majorana fermions coupled to a static $Z_2$ gauge field in the extended Hilbert space. From this, he explicitly showed that the ground state of the model is a gapless $Z_2$ QSL, while the fractionalized excitations are gapless (or gapped for anisotropic exchange couplings) Majorana fermions and gapped $Z_2$ gauge fluxes.\cite{Kitaev2006}

Besides the Kitaev representation with four Majorana fermions,\cite{Kitaev2006} two other types of Majorana representation of spin are known. The first one was introduced by Tsvelik. \cite{Tsvelik1992, tsvelikbook} This representation contains three Majorana fermions transforming under SO(3), thus we will call it  the {\it SO(3) Majorana representation}. The SO(3) Majorana representation has a significant advantage over the Kitaev representation that no unphysical states or extension of the Hilbert space is involved.\cite{Shnirman2003,Mao2003} Using this representation, various spin models have been studied on the mean-field level including one-dimensional spin chain \cite{Shastry1997,Nersesyan2011}  and two-dimensional triangular lattices.\cite{Biswas2011,Herfurth2013} The third type of Majorana fermion representation was introduced by Chen {\it et al.} in their study of QSLs realized on a two-dimensional square lattice using the projective symmetry group (PSG).\cite{Gang2012} We will call it the {\it SO(4) chiral Majorana representation} for reasons that will become clear later. 

Knowing these three Majorana representations, this paper addresses the following natural questions: First, what is the connection between these three types of Majorana representation of spin? Second, is there any relationship between these Majorana representations and the complex fermion representation? \cite{Marston1989,Lee2006} Third, can one use different types of Majorana representations to study the same system obtaining similar results but also for gaining new insights?   

Here, in response to the first question, we show that the three different Majorana fermion representations of spin can be motivated as a whole from the Dirac spinor representation of the SO(4) group which is due to the fact that the Majorana fermions satisfy the same Clifford algebra as the Gamma matrices introduced in the representation of the Lorentz group. 

For the second question, we show that there is a close correspondence between the SO(4) chiral Majorana representation and the complex fermion representation, as has been discussed in previous works. \cite{Kitaev2006, Burnell2011,Shnirman2003,Biswas2011,Gang2012} In particular, the Hilbert space of the two types of representations involving four Majorana fermions per sites (the Kitaev representation and the SO(4) chiral representation) have a four-dimensional Hilbert space per site and require a local projection onto the two-dimensional spin Hilbert space in analogy to the {\it chiral projection}. Such Hilbert space can be mapped to the Hilbert space of the complex fermion representation. On the other hand, the SO(3) Majorana representation has three Majorana fermions defined on each sites, its Hilbert space involves multiple copies of the physical space with no unphysical degree of freedom. Upon choosing a suitable pairing of the sites, it was shown by Biswas {\it et al}~\cite{Biswas2011} that the physical spin space is faithfully represented. 

To answer the third question, we study the exactly soluble Kitaev model. We show that the SO(3) Majorana representation gives an alternative exact solution. We focus on how to get the physical states of the model and give an explicit formula which involves $Z_{2}$ gauge transformations of the states. These results can be directly compared with  Kitaev's original solution providing an answer to the third question. Finally, the solution of the Kitaev model takes the form of a lattice $Z_{2}$ gauge theory interacting with complex fermions. We extended it to a more general lattice gauge theory with plaquette terms and sketch its ground state phase diagram. 

The rest of the paper is organized as follows. In Section \ref{secthreetypes}, we introduce the three types of Majorana representation of spin and discuss their Majorana Hilbert space and the relationship to the spin Hilbert space. In Section \ref{secspinorrep}, we first review the spinor representation of the Lorentz group and the SO(4) group and argue that the three types of Majorana representation of spin have close mathematical relationship with the Dirac spinor representation of the SO(4) group. From that we relate the projection operator in the Kitaev representation to the chirality projection in the spinor representation. With this understanding of the Majorana representations, we emphasize that the SO(3) Majorana representation has an advantage of faithfully reproducing the physical Hilbert space with proper pairing of spin sites. \cite{Biswas2011} We then illustrate this point by presenting another solution to the Kitaev model using this representation, this is given in Section \ref{seckitaevmodel}. We show that with this representation a generalized Kitaev model can be transformed into a $Z_{2}$ gauge theory of complex fermions on a diamond lattice (see Fig. \ref{kitaevlattice}). We show how to obtain the physical states explicitly and sketch the possible phases realized in a generalized Kitaev model. A discussion and summary of the results is given in Section \ref{secconclusion}. 

\section{Three Types of Majorana Representation of Spin} \label{secthreetypes}

The three types of Majorana representations of spin-1/2  degrees  of freedom fall into two categories:  the SO(3) Majorana representation uses three Majorana fermions to represent a single spin operator  while the Kitaev representation and the SO(4) chiral Majorana representation use four Majorana fermions to represent a single spin operator. As a basic criteria these representations need to satisfy the spin-1/2 algebra, namely,
\begin{equation}
\label{spinrelation}
\sigma^{\alpha}\sigma^{\beta}=\delta^{\alpha\beta}+i\epsilon^{\alpha\beta\gamma}\sigma^{\gamma}, \qquad \alpha,\beta,\gamma=x,y,z.
\end{equation}
Below we  discuss the details of the three types of representations starting with the SO(3) Majorana representation. 

\subsection{SO(3) Majorana representation}\label{subsecso3}

For the SO(3) Majorana representation, we first introduce for each spin operator $\boldsymbol{\sigma}_{i}$ three Majorana fermion operators $\eta_{i}^{x}$, $\eta_{i}^{y}$ and $\eta_{i}^{z}$, transforming under the fundamental representation of SO(3). Majorana fermions are real, $\eta_{i}^{\alpha}=\eta_{i}^{\alpha\dagger}$, and they satisfy the {\it Clifford algebra} $\{\eta^{\alpha}_{i},\eta^{\beta}_{j}\}=2\delta^{\alpha\beta}\delta_{ij}$ when $\alpha\neq\beta$ or $i\neq j$, it means they are anticommuting fermions, and when $\alpha=\beta, i=j$, it means $(\eta^{\alpha}_{i})^2=1$ (from now on we use $\alpha,\beta$ to label indices $x, y, z$). The SO(3) Majorana representation is then given by \cite{Tsvelik1992, tsvelikbook}
\begin{equation}
\label{so3majoranarepresentation}
\sigma^{x}_{i}=-i\eta^{y}_{i}\eta^{z}_{i}, \qquad \sigma^{y}_{i}=-i\eta^{z}_{i}\eta^{x}_{i}, \qquad \sigma^{z}_{i}=-i\eta^{x}_{i}\eta^{y}_{i}, 
\end{equation}
or  in a more compact form,
\begin{equation}
\label{so3majoranarep}
\sigma^{\alpha}_{i}=-\frac{i}{2}\epsilon^{\alpha\beta\gamma}\eta^{\beta}_{i}\eta^{\gamma}_{i}. \qquad \alpha,\beta,\gamma=x,y,z .
\end{equation}
Since the Majorana fermions have the aforementioned properties, it is clear that this representation (\ref{so3majoranarepresentation}) satisfies the spin relation (\ref{spinrelation}) automatically. 

To make further progress, we define a new operator $\tau_{i}$ for each spin as\cite{Shnirman2003,Mao2003}
\begin{equation}
\label{tauoperator}
\tau_{i}=-i\eta^{x}_{i}\eta^{y}_{i}\eta^{z}_{i}.
\end{equation}
It can be shown that this operator has the following properties: (i) it transforms as singlet under SO(3); (ii) it commutes with the three Majorana fermions on the same site: $[\tau_{i}, \eta_{i}^{\alpha}]=0$; (iii) it anticommutes with three Majorana fermions $\{\tau_{i}, \eta_{j}^{\alpha}\}=0$ for different sites $i$ and $j$, furthermore, due to the fact that spins are bilinear in $\eta$, we have $[\tau_{i}, \sigma_{j}^{\alpha}]=0$ for different sites $i$ and $j$. Combining properties (ii) and (iii) and noting that spins are bilinear in $\eta$-operators, we find that the operator $\tau$ commutes with all spin operators, \cite{Shnirman2003}
\begin{equation}
[\tau_{i},\sigma_{j}^{\alpha}]=0,\qquad \alpha=x,y,z
\end{equation}
and thus it commutes with any  spin Hamiltonian. So the $\tau$ operator is a fermionic constant for each site.   Using both $\eta$ and $\tau$ operators we can now write the spin operators as
\begin{equation}
\label{so3majoranarep2}
\sigma_{i}^{x}=\tau_{i}\eta_{i}^{x}, \qquad \sigma_{i}^{y}=\tau_{i}\eta_{i}^{y}, \qquad \sigma_{i}^{z}=\tau_{i}\eta_{i}^{z}.
\end{equation}
It is important to note that the representation (\ref{so3majoranarep}) has a local $Z_{2}$ gauge redundancy in the Majorana fermion space, i.e. the spin operator is invariant under local transformation $\eta_{i}^{\alpha}\rightarrow \epsilon_{i}\eta_{i}^{\alpha}$ with $\epsilon_{i}=\pm 1$. Under such gauge transformation, the $\tau$ operator transforms as $\tau_{i}\rightarrow\epsilon_{i}\tau_{i}$.

Before we move on, it is important to understand the Hilbert space of the SO(3) Majorana representation and how to map from the Hilbert space of three Majorana fermions to the spin Hilbert space. Each Majorana fermion has nominally a Hilbert space dimension of $\sqrt{2}$, therefore, one can introduce an extra Majorana fermion for each spin followed by some projection.\cite{Shnirman2003} An alternative way, which we are going to follow, is to pair up the spins and define the Hilbert space in a non-local way. \cite{Biswas2011} In particular, since for $N$ spins there are $3N$ Majorana fermions, the dimension of the Hilbert space of the Majorana fermions is $2^{3N/2}$, which is $2^{N/2}$ times larger than the dimension of the original spin Hilbert space. To gain a one-to-one correspondence, we first group the $N$ spins into $\frac{N}{2}$ pairs,\cite{Biswas2011} and then for each pair $\langle ij\rangle$ we define the product operator $\tau_{i}\tau_{j}$, with $\tau$-operators defined in Eq.(\ref{tauoperator}). Since these product operators for every pair commute with each other, we can choose the eigenvalues of $\tau_{i}\tau_{j}$ to be either $+i$ or $-i$ for every pair. Of course, the number of choices is $2^{N/2}$ for $\frac{N}{2}$ pairs. For each choice, one can prove that the Hilbert space has a one-to-one correspondence with the spin Hilbert space,\cite{Biswas2011} and that the Majorana Hilbert space is just $2^{N /2}$ copies of the original spin Hilbert space. Here, the $Z_{2}$ gauge symmetry plays an important role. We have the freedom of flipping the sign of Majorana operators on each site, generating $2^{N}$ gauge copies, but the flipping of the sign of each of the two sites belonging to the same pair leaves the sign of the $\tau_{i}\tau_{j}$ operator of the pair unchanged. So the net number of  gauge copies is actually $2^{N/2}$, which means that the redundancy of the dimension of the Majorana Hilbert space is directly related to the $Z_{2}$ gauge symmetry of the theory.

\subsection{Kitaev representation}\label{subsecsokitaev}
For the Kitaev representation \cite{Kitaev2006} one  introduces four Majorana fermions for each spin operator. For clarity, we follow the notation of the previous section and write them as $\eta_{i}^{t},\eta_{i}^{x},\eta_{i}^{y},\eta_{i}^{z}$. The four Majorana fermions transform in the fundamental representation of SO(4) and they satisfy the {\it Clifford algebra} (from now on we use $\mu,\nu$ to label index $t, x, y, z$)
\begin{equation}
\label{majoranaclifford}
\{\eta_{i}^{\mu},\eta_{j}^{\nu}\}=2\delta^{\mu\nu}\delta_{ij}, \qquad \mu,\nu=t,x,y,z.
\end{equation}  
In the Kitaev representation,\cite{Kitaev2006} the spin operator is given by a product of two Majorana fermions as
\begin{equation}
\label{kitaevrep}
\sigma_{i}^{\alpha}=i\eta^{t}_{i}\eta_{i}^{\alpha} \qquad \alpha=x,y,z.
\end{equation}
This Majorana representation of spins is overcomplete. In particular, the dimension of the Hilbert space of the Majorana fermions is 4 for each site, while the dimension is only 2 for the physical spin space. This means that the Majorana Hilbert space contains both physical and unphysical states. Thus it is necessary to project out the unphysical part. It was proven by Kitaev that the representation (\ref{kitaevrep}) satisfies the spin relation (\ref{spinrelation}) under the constraint that \cite{Kitaev2006}
\begin{equation}
\label{so4constraint}
D_{i}=\eta_{i}^{t}\eta_{i}^{x}\eta_{i}^{y}\eta_{i}^{z}=1 \qquad \text{for every physical state}.
\end{equation} 

To enforce the constraint (\ref{so4constraint}) we define the {\it chirality projection operator} as
\begin{equation}
\label{chiprojector}
P_{i,L}=\frac{1+D_{i}}{2},
\end{equation} 
for which the meaning of the subscript index ``{\it L}" will become clear shortly. The physical states can now be written as $\prod_{i}P_{i,L}|\psi\rangle$, where $|\psi\rangle$ is a state in the extended Hilbert space of Majorana fermions. This procedure leads us to the definition of the third type of Majorana representation.\cite{Gang2012}

\subsection{SO(4) chiral Majorana representation}\label{subsecso4}

Combining the Kitaev representation (\ref{kitaevrep}) and the chirality projector (\ref{chiprojector}), we obtain another type of Majorana representation of spin, 
\begin{equation}
\label{so4chiralrep}
\sigma_{i}^{\alpha}=P_{i,L}(i\eta^{t}_{i}\eta_{i}^{\alpha})=\frac{i}{2}(\eta^{t}_{i}\eta^{\alpha}_{i}-\frac{1}{2}\epsilon^{\alpha\beta\gamma}\eta^{\beta}_{i}\eta^{\gamma}_{i}), 
\end{equation} 
where $\alpha,\beta,\gamma=x,y,z$.
Written out explicitly, Eq. (\ref{so4chiralrep}) reads
\begin{eqnarray}
\begin{aligned}
\label{so4chiral2}
&\sigma^{x}_{i}=\frac{i}{2}(\eta^{t}_{i}\eta^{x}_{i}-\eta^{y}_{i}\eta^{z}_{i}),\\
&\sigma^{y}_{i}=\frac{i}{2}(\eta^{t}_{i}\eta^{y}_{i}-\eta^{z}_{i}\eta^{x}_{i}),\\ &\sigma^{z}_{i}=\frac{i}{2}(\eta^{t}_{i}\eta^{z}_{i}-\eta^{x}_{i}\eta^{y}_{i}).
\end{aligned}
\end{eqnarray}
It can be shown that this representation satisfies the spin relation (\ref{spinrelation}) with the constraint (\ref{so4constraint}). We will call this representation (\ref{so4chiralrep}) the SO(4) chiral Majorana representation. 

The spin operator defined in the SO(4) chiral representation (\ref{so4chiralrep}) satisfies $[\sigma_{i}^{\alpha},P_{i,L}]=0$, which means that the projection onto the physical space has to be done only once, in other words, any spin term acting on a physical state gives a physical state. However, to achieve this, we end up with a representation (\ref{so4chiral2}) much more complex than the simple Kitaev representation (\ref{kitaevrep}). Moreover, we note that in Ref.\onlinecite{Gang2012} the SO(4) chiral Majorana representation was introduced in a  different way, there, the representation was obtained complementary to the complex fermion representation using the PSG method.\cite{Gang2012} Although the SO(4) chiral Majorana representation can be seen as a direct generalization of the Kitaev representation, we will treat it as another type of Majorana representation. To see the reason, we move on to discuss the Hilbert space of the four Majorana fermions defined in both Kitaev representation and SO(4) chiral Majorana representation. From that, we shall see that the SO(4) chiral Majorana representation has a direct correspondence to the familiar complex fermion one. \cite{Gang2012}   

Using representation (\ref{so4chiral2}), we can obtain the spin raising and lowering operators 
\begin{eqnarray}\label{raisinglowering}
\begin{aligned}
&\sigma^{+}_{i}=\frac{1}{4}(\eta^{z}_{i}+i\eta^{t}_{i})(\eta^{x}_{i}+i\eta^{y}_{i}), \\
&\sigma^{-}_{i}=\frac{1}{4}(\eta^{x}_{i}-i\eta^{y}_{i})(\eta^{z}_{i}-i\eta^{t}_{i}).
\end{aligned}
\end{eqnarray}
From this we define two complex fermions $f_{i}=\frac{1}{2}(\eta^{z}_{i}-i\eta^{t}_{i}),f^{\dagger}_{i}=\frac{1}{2}(\eta^{z}_{i}+i\eta^{t}_{i})$ and $g_{i}=\frac{1}{2}(\eta^{x}_{i}-i\eta^{y}_{i}),g^{\dagger}_{i}=\frac{1}{2}(\eta^{x}_{i}+i\eta^{y}_{i})$. For reasons which will become clear later, we may just label $f_{i}=f_{i,\uparrow}$ and $g_{i}=f_{i,\downarrow}^{\dagger}$, then we have the following relations
\begin{eqnarray}\label{fupfdown}
\begin{aligned}
&f_{i,\uparrow}=\frac{1}{2}(\eta^{z}_{i}-i\eta^{t}_{i}),\,\,f_{i,\uparrow}^\dagger=\frac{1}{2}(\eta^{z}_{i}+i\eta^{t}_{i}),\\
& f_{i,\downarrow}=\frac{1}{2}(\eta^{x}_{i}+i\eta^{y}_{i}),\,\, f_{i,\downarrow}^\dagger=\frac{1}{2}(\eta^{x}_{i}-i\eta^{y}_{i}).
\end{aligned}
\end{eqnarray}
Conversely, the Majorana fermions can be expressed in terms of these complex fermions as \cite{Gang2012}
\begin{eqnarray}
\begin{aligned}
\label{convers}
&\eta^{t}_{i}=i(f_{i,\uparrow}-f_{i,\uparrow}^{\dagger}),\\
& \eta^{x}_{i}=f_{i,\downarrow}+f_{i,\downarrow}^{\dagger},
\\ 
&\eta^{y}_{i}=i(f_{i,\downarrow}^{\dagger}-f_{i,\downarrow}), \\&
\eta^{z}_{i}=f_{i,\uparrow}+f_{i,\uparrow}^{\dagger}.
\end{aligned}
\end{eqnarray}
Using Eq.(\ref{convers}), we can transform the spin raising and lowering operators expressed under the SO(4) chiral Majorana representation (\ref{raisinglowering}) into 
\begin{eqnarray}
\begin{aligned}
\label{complexslaverep}
&\sigma^{+}_{i}=f_{i,\uparrow}^{\dagger}f_{i,\downarrow},
\,\, \sigma^{-}_{i}=f_{i,\downarrow}^{\dagger}f_{i,\uparrow},
\\ 
&\sigma^{z}_{i}=f_{i,\uparrow}^{\dagger}f_{i,\uparrow}-f_{i,\downarrow}^{\dagger}f_{i,\downarrow}.
\end{aligned}
\end{eqnarray}
  Eq.(\ref{complexslaverep})  shows that we have recovered the familiar complex fermion slave particle representation 
$
\sigma^{\alpha}=f_{\beta}^{\dagger}(\tilde{\sigma}^{\alpha})_{\beta\gamma}f_{\gamma},
$
with $\tilde{\sigma}$ being the Pauli matrices and $f_{\alpha}$ being the complex slave particles called spinons. From  Eq. (\ref{convers}) we also have the relation $\eta^{t}_{i}\eta^{x}_{i}\eta^{y}_{i}\eta^{z}_{i}=-(1-2n_{i,\uparrow})(1-2n_{i,\downarrow})$ ($n_{i,\uparrow}$ and $n_{i,\downarrow}$ are the number of complex fermion on each site), from which we can clearly see that the constraint (\ref{so4constraint}) is equivalent to the constraint in the complex fermion representation that each site has only one fermion, i.e. $f_{i,\uparrow}^{\dagger}f_{i,\uparrow}+f_{i,\downarrow}^{\dagger}f_{i,\downarrow}=1$. 

From the mapping defined in  Eqs.(\ref{fupfdown}) and (\ref{convers}) it is clear that the Hilbert space of the four Majorana fermions introduced in both the Kitaev representation and the chiral Majorana representation can be mapped to the Hilbert space of the familiar complex fermions (\ref{complexslaverep}). Moreover, the constraint (\ref{so4constraint}) for the Majorana representations and the familiar one-fermion-per-site constraint \cite{Lee2006}  for the complex fermion representation are also mapped into each other. \cite{Burnell2011, Gang2012} Therefore, the SO(4) chiral Majorana representation can be seen as exactly equivalent to the complex fermion representation and it is defined in the same Hilbert space as the Kitaev representation, thus it acts like a ``bridge" between the two seemingly unrelated represetations.

In the following, to see the connection between the three types of Majorana representations, we will explore the Clifford algebra of the Majorana fermions in detail. It allows us to link all the representations to the spinor representation of the SO(4) group and the Lorentz group.

\section{Spinor representation of the SO(4) group and connection between Majorana representations} \label{secspinorrep}

\subsection{Representation of the Lorentz group}

In order to see the relationship between the three types of Majorana representations of spin introduced above and the spinor representation of SO(4), we recapitulate the representation of the Lorentz group SO(1,3), which has almost identical structure but terminology more familiar. Strictly speaking the Lorentz group is not exactly the group SO(1,3) but in this paper, we neglect such difference as long as no confusion is caused. We will follow and use the notations of Ref. \onlinecite{schwartzbook} and define the space-time metric as $g_{\mu\nu}=\text{diag}(1,-1,-1,-1)$. In the 4-vector representation, the Lorentz transformation is written as  $\Lambda_{v}=\exp(i\theta_{\mu\nu}V^{\mu\nu})$, in which $\theta_{\mu\nu}$ are the parameters charactering the transformation, and $V^{\mu\nu}$ are the generators of the Lorentz algebra. Due to symmetry, only six parameters of $\theta_{\mu\nu}$ are independent, three of them characterize the space rotation and the other three characterize the Lorentz boost. Taking this into account, the Lorentz elements in the 4-vector basis is also written as $\Lambda_{v}=\exp(i\theta_{i}J_{i}+i\beta_{i}K_{i})$, in which $J_{i}$ are the generators of  rotation and $K_{i}$ are the generators of  boost.

The Lie algebra of the Lorentz group $\mathcal{L}(\text{SO(1,3)})=\mathrm{so}(1,3)$ can be decomposed into two commuting subalgebra $\mathrm{so}(1,3)=\mathrm{su}(2)\oplus\mathrm{su}(2)$, with the generators defined as $J_{i}^{+}\equiv\frac{1}{2}(J_{i}+iK_{i}), \qquad J_{i}^{-}\equiv\frac{1}{2}(J_{i}-iK_{i})$. They satisfy two separate SU(2) Lie algebra,
\begin{eqnarray}
\begin{aligned}
&[J_{i}^{+},J_{j}^{+}]=i\epsilon_{ijk}J_{k}^{+},\\  &[J_{i}^{-},J_{j}^{-}]=i\epsilon_{ijk}J_{k}^{-},\\ &[J_{i}^{+},J_{j}^{-}]=0.
\end{aligned}
\end{eqnarray}
Since the representations for a single SU(2) group are the angular momentum eigenstates, we can label the representation of the Lorentz group as $(j_{1},j_{2})$, corresponding to the two su(2) subalgebra. The most fundamental but non-trivial representation is $(\frac{1}{2},0)$ and $(0,\frac{1}{2})$, these are the {\it Weyl spinor representations}. The $(0,\frac{1}{2})$ is called right-handed Weyl spinor, it transforms as $\psi_{R}\rightarrow e^{\frac{1}{2}(i\theta_{i}\sigma_{i}+\beta_{i}\sigma_{i})}\psi_{R}$ under the Lorentz group. The $(\frac{1}{2},0)$ is called left-handed Weyl spinor and it transforms as $\psi_{L}\rightarrow e^{\frac{1}{2}(i\theta_{i}\sigma_{i}-\beta_{i}\sigma_{i})}\psi_{L}$ under the Lorentz group. Again, we have used $\theta_{i}$ and $\beta_{i}$ to characterize the space rotation and the boost.

The Weyl spinors are irreducible representation of the Lorentz group and they are the building blocks of the Dirac spinors, which are of fundamental importance to elementary particle physics \cite{schwartzbook}. Dirac spinors live in the representation $(\frac{1}{2},0)\oplus(0,\frac{1}{2})$, and normally it is written in terms of Weyl spinors as $\psi=(\psi_{L},\psi_{R})^{T}$. In order to study the Lorentz group in the Dirac spinor representation, it is necessary to introduce the $\gamma$-matrices, which were first used in the Dirac equations of relativistic quantum mechanics. \cite{schwartzbook} The $\gamma$-matrices are four $4\times 4$ matrices satisfying the {\it Clifford algebra}
\begin{equation}
\label{gammacliffordalgebra}
\{\gamma^{\mu},\gamma^{\nu}\}=2g^{\mu\nu}, \qquad \mu,\nu=0,1,2,3.
\end{equation}
Using $\gamma$-matrices we can introduce the generators of the Lorentz group in the Dirac spinor representation as 
\begin{equation}
\label{spinorgenerator}
S^{\mu\nu}=\frac{i}{4}[\gamma^{\mu},\gamma^{\nu}].
\end{equation}
They satisfy the Lorentz algebra 
\begin{equation}
\label{lorentzalgebra}
[S^{\mu\nu},S^{\rho\sigma}]=i(g^{\nu\rho}S^{\mu\sigma}-g^{\mu\rho}S^{\nu\sigma}-g^{\nu\sigma}S^{\mu\rho}+g^{\mu\sigma}S^{\nu\rho}).
\end{equation}
A general Lorentz transformation in the Dirac spinor representation is written as $\Lambda_{s}=\exp(i\theta_{\mu\nu}S^{\mu\nu})$. It is often useful to project a Dirac spinor to its left-handed or right-handed Weyl spinor component (called chirality projection). To do so, it is essential to introduce another matrix $\gamma^{5}$ defined as $\gamma^{5}\equiv i\gamma^{0}\gamma^{1}\gamma^{2}\gamma^{3}$. The chirality projectors are thus given by $P_{R}=\frac{1+\gamma^{5}}{2}$ and $P_{L}=\frac{1-\gamma^{5}}{2}$. The $\gamma^{5}$ matrix satisfies $\{\gamma^{5},\gamma^{\mu}\}=0$, and $(\gamma^{5})^{2}=1$.

\subsection{Three Types of Majorana Representation of Spin and the Spinor Representation of SO(4)}\label{secmajoranaspinor}

The Clifford algebra satisfied by the four Majorana fermions for each spin (\ref{majoranaclifford}) differs from the Clifford algebra of $\gamma$ matrices (\ref{gammacliffordalgebra}) only in the metric. Such difference in the metric is the source of the marginal difference of the group SO(4) and SO(1,3). Indeed, if we were to redefine the Majorana fermion $\eta^{t}\rightarrow i\eta^{t}$, the Clifford algebra of the Majorana fermions and the Clifford algebra of $\gamma$ matrices would be the same. However for simplicity, there is no need for such redefinition.

Next we follow the steps of building the Dirac spinor representation of Lorentz group and define the objects 
\begin{equation}
\label{definitionofS}
\mathcal{S}^{\mu\nu}\sim i[\eta^{\mu},\eta^{\nu}]
\end{equation}
according to  Eq. (\ref{spinorgenerator}). Such object $\mathcal{S}^{\mu\nu}$ satisfies the following algebra
\begin{equation}
[\mathcal{S}^{\mu\nu},\mathcal{S}^{\rho\sigma}]=i(\delta^{\nu\rho}\mathcal{S}^{\mu\sigma}-\delta^{\mu\rho}\mathcal{S}^{\nu\sigma}-\delta^{\nu\sigma}\mathcal{S}^{\mu\rho}+\delta^{\mu\sigma}\mathcal{S}^{\nu\rho})
\end{equation}
and thus are the generators of the group SO(4) in some {\it Dirac-spinor-like representation}. The vector space this representation is acting on is the Hilbert space of the four Majorana fermion, which will be explored in more detail below. Although the four space-time directions $t, x, y, z$ have identical footing in SO(4), it is more convenient for us to keep the terminology of the Lorentz group, which puts time direction in a special position. 

From the definition Eq.(\ref{definitionofS}), one can find that the components $\mathcal{S}^{0\alpha}\sim i[\eta^{t},\eta^{\alpha}]\sim i\eta^{t}\eta^{\alpha}$ give exactly the Kitaev representation (\ref{kitaevrep}). In the Lorentz group terminology, $\mathcal{S}^{0\alpha}$ correspond to the generators of the ``Lorentz boosts". On the other hand, the ``space rotation" components, $\mathcal{S}^{\alpha\beta}\sim i[\eta^{\alpha},\eta^{\beta}]\sim i\eta^{\alpha}\eta^{\beta}$, gives the SO(3) Majorana representation (\ref{so3majoranarep}), if again we keep the same terminology. The Dirac spinor representation of SO(4) can be decomposed into the left-handed and the right-handed Weyl spinor representation, just like the Lorentz group SO(1,3): $\text{SO(4)}=\text{SU(2)}_{L}\times\text{SU(2)}_{R}$. Therefore we can still define the {\it chirality projection} operator in the Dirac-spinor-like representation 
\begin{equation}
P_{L}=\frac{1+\eta^{t}\eta^{x}\eta^{y}\eta^{z}}{2}
\end{equation}
as in the Lorentz group. This leads us to intepret the projection defined in  Eq. (\ref{chiprojector}) as the projection to the left-chirality in the Dirac-spinor-like representation. Now the meaning of the definitions given in the previous section should become clear. 

Therefore, we find that there is a clear connection between the three types of Majorana representation of spin and the Dirac-spinor-like representation of SO(4). In particular, we find that in a loose sense, the SO(3) Majorana representation (\ref{so3majoranarep}) corresponds to the ``space rotation" SO(3) subgroup of SO(4), the generators of which give the spin relation automatically. The Kitaev representation (\ref{kitaevrep}), corresponding to the ``Lorentz boost" part of SO(4), does not have the desired SO(3) structure. But once we project all the {\it states} to one of the chirality (say left), the Kitaev representation will satisfy the spin relation (\ref{spinrelation}). On the other hand, the SO(4) chiral Majorana representation (\ref{so4chiralrep}), as the projected Kitaev representation to the left-chirality, is the generators of $\text{SU(2)}_{L}$ in the Dirac-spinor-like representation of SO(4). It has the desired spin commutator, but the constaint (\ref{so4constraint}) is still needed to ensure the normalization $(\sigma^{\alpha})^{2}=1$.   

Now we examine the Hilbert space of the four Majorana fermions $\eta^{t},\eta^{x},\eta^{y},\eta^{z}$, which is also the vector space the Dirac-spinor-like representation of SO(4) is acting on. From the discussion in the previous section, we see that this Hilbert space is 4-dimensional and is the same as the Hilbert space of the two complex fermions $f_{\uparrow}$, $f_{\downarrow}$, which has basis vectors $(|\uparrow\rangle,|\downarrow\rangle),(|0\rangle,|\uparrow\downarrow\rangle)$. This means that there is a one-to-one mapping between SO(4) Dirac spinor space and the Hilbert space of 2 complex fermions. A generalization of this statement for the mapping between the Dirac spinor space of SO($2N$) and the Hilbert space of $N$ complex fermions where $N$ is an integer has been proposed in high energy physics \cite{Mohapatra80,Wilczek82} and later used in the description of non-Abelian anyons. \cite{Chetan1996,Ahlbrecht2009} Readers can find detailed mathematical description of this mapping in these references. For our purpose of describing spin in four space-time dimensions, the SO(4) group is sufficient. On the other hand, generalizations to higher SO($2n$) where $n>2$ will involve spinors in higher dimensional space. In Appendix \ref{appendixmapping}, we give a more detailed discussion of the mapping for the SO(4) group and other related issues. 

Although the mathematical discussion in this section is far from rigorous, it serves as a tool to help us think about the representations we have so far and to compare them. From the discussion above, we see that the complex fermion representation might not be the most fundamental representation of spin as it seems; rather it can be understood as a part of the bigger class of representation, the Majorana fermion representation. Among the three types of Majorana fermion representation at hand, the SO(3) Majorana representation is special because it only has three Majorana fermions instead of four. The Hilbert space is thus not complete and not well-defined for each site. A pairing of sites is required for a proper definition of the Hilbert space. \cite{Biswas2011} In a sense, the Hilbert space defined in this way no longer possesses the properties discussed above and requires further considerations.  Here, we will not go in this direction. Instead, to answer the third question raised in the introduction (Sec. \ref{introduction}) we explicitly show how to solve the seminal Kitaev model via the SO(3) Majorana representation.

\section{Physical solution of the Kitaev model using the SO(3) Majorana representation} \label{seckitaevmodel}

\subsection{The Kitaev Model}

The Kitaev model is a two-dimensional exactly solvable model of spin-1/2 degrees of freedom defined on the honeycomb lattice.\cite{Kitaev2006} The bonds of the honeycomb lattice are categorized into three types which are labelled by $x$, $y$, and $z$, and denoted by $\langle ij\rangle_{a}$ where $a=x,y,z$. Spins interact with its nearest neighbours via an anisotropic Ising-like interaction. In particular, on each type of bond only the corresponding spin components are interacting (see Fig \ref{kitaevlattice}). The Hamiltonian of the model is given by,
\begin{equation}
\label{kitaevhamiltonian}
\mathcal{H}=J_{x}\sum_{\langle ij\rangle_{x}}\sigma_{i}^{x}\sigma_{j}^{x}+J_{y}\sum_{\langle ij\rangle_{y}}\sigma_{i}^{y}\sigma_{j}^{y}+J_{z}\sum_{\langle ij\rangle_{z}}\sigma_{i}^{z}\sigma_{j}^{z},
\end{equation}
where $J_{x}, J_{y}$ and $J_{z}$ are the Ising coupling strengths on the $x$, $y$, and $z$ bonds, respectively. 

The Hamiltonian (\ref{kitaevhamiltonian}) can be solved exactly using the Kitaev representation of spins (\ref{kitaevrep}).\cite{Kitaev2006}  The resulting theory is a $Z_{2}$ quantum spin liquid, which is either gapped or gapless depending on the values of $J_{x}, J_{y}$ and $J_{z}$. As we discussed above, since the representation (\ref{kitaevrep}) is defined in the extended Hilbert space, every calculation of the model using this representation should be projected onto the physical space in each step. Such projection based on the chirality constraint (\ref{so4constraint}) has been discussed in Refs. \onlinecite{Kitaev2006,Pedrocchi2011,zschocke2015}, where it has been explicitly shown for small system sizes that quantities computed in the extended Hilbert space can be substantially different from the ones calculated in the physical space.\cite{Pedrocchi2011} In general the projection is difficult to implement. Therefore, more systematic ways of obtaining the physical solution are desired to better understand the model and to make reliable predictions. Some previous works have already made progress in this direction. In paticular, it has been shown that it is possible to achieve a solution of the model without using the slave-particle representation of spin, e.g., using Jordan-Wigner transformation.\cite{ChenNussinov2008,Kells2009,Dora2018} Here, we will stick with the slave-particle approach and show that it is possible to obtain the solution of the Kitaev model using the SO(3) Majorana representation of spin instead of the Kitaev representation. The advantage is that without the extension of the Hilbert space, our alternative solution is automatically physical.

\subsection{Solution of the Kitaev model via the SO(3) Majorana representation}\label{secsolutionofkitaev}

In this section we show how the solution of the Kitaev model (\ref{kitaevhamiltonian}) can be obtained using the SO(3) Majorana representation. As described in Sec.\ref{subsecso3},   we first introduce three Majorana fermions for each spin and label them as $\eta_{i}^{x},\eta_{i}^{y},\eta_{i}^{z}$. We then rewrite the Hamiltonian (\ref{kitaevhamiltonian}) using representations (\ref{so3majoranarepresentation}) and (\ref{so3majoranarep2}) as follows
\begin{widetext}
\begin{eqnarray}
\label{Ham1}
\begin{aligned}
\mathcal{H}=&\sum_{\langle ij\rangle_{x}}J_{x}(-i\eta_{i}^{y}\eta_{i}^{z})(-i\eta_{j}^{y}\eta_{j}^{z})+\sum_{\langle ij\rangle_{y}}J_{y}(-i\eta_{i}^{z}\eta_{i}^{x})(-i\eta_{j}^{z}\eta_{j}^{x})+\sum_{\langle ij\rangle_{z}}J_{z}(\tau_{i}\eta_{i}^{z})(\tau_{j}\eta_{j}^{z})\\
=&\sum_{\langle ij\rangle_{x}}J_{x}(\eta_{i}^{y}\eta_{j}^{y})\eta_{i}^{z}\eta_{j}^{z}+\sum_{\langle ij\rangle_{y}}J_{y}(\eta_{i}^{x}\eta_{j}^{x})\eta_{i}^{z}\eta_{j}^{z}+\sum_{\langle ij\rangle_{z}}J_{z}(-\tau_{i}\tau_{j})\eta_{i}^{z}\eta_{j}^{z}.
\end{aligned}
\end{eqnarray}
\end{widetext}
In particular, for all the z-bonds we use Eq. (\ref{so3majoranarep2}) to represent spins while for x-bonds and y-bonds we apply representation (\ref{so3majoranarepresentation}).

For the second step, we group the two sites belonging to every $z$-bond together and require that 
\begin{equation}
\label{conditionofpair}
\tau_{i}\tau_{j}=-i, \text{ for each } \langle ij\rangle_{z}  
\end{equation}
with $i$ belonging to the A sublattice. As discussed above, such pairing of honeycomb lattice sites and condition (\ref{conditionofpair}) eliminate all the extra degrees of freedom and the Hilbert space for $\eta_{i}^{x},\eta_{i}^{y},\eta_{i}^{z}$ is now the same as the spin space. With the condition (\ref{conditionofpair}), the Hamiltonian is transformed into
\begin{equation}
\label{hprime}
\mathcal{H}^{'}=\sum_{\langle ij\rangle_{x}}J_{x}(\eta_{i}^{y}\eta_{j}^{y})\eta_{i}^{z}\eta_{j}^{z}+\sum_{\langle ij\rangle_{y}}J_{y}(\eta_{i}^{x}\eta_{j}^{x})\eta_{i}^{z}\eta_{j}^{z}+\sum_{\langle ij\rangle_{z}}iJ_{z}\eta_{i}^{z}\eta_{j}^{z}.
\end{equation}
Note that we have $[\eta_{i}^{y}\eta_{j}^{y},\mathcal{H}^{'}]=0$ for $\langle ij\rangle_{x}$ and $[\eta_{i}^{x}\eta_{j}^{x},\mathcal{H}^{'}]=0$ for $\langle ij\rangle_{y}$, therefore we are free to pick up eigenvalues to solve the transformed Hamiltonian $\mathcal{H}^{'}$, one can choose $\eta_{i}^{y}\eta_{j}^{y}=\pm i$ for  $\langle ij\rangle_{x}$ and  $\eta_{i}^{x}\eta_{j}^{x}=\pm i$ for $\langle ij\rangle_{y}$. With this, the Hamiltonian is finally transformed to a free hopping Hamiltonian for $\eta^{z}$ Majorana fermions,
\begin{equation}
\label{hdoubleprime}
\mathcal{H}^{''}=\sum_{\langle ij\rangle_{x}}(\pm i)J_{x}\eta_{i}^{z}\eta_{j}^{z}+\sum_{\langle ij\rangle_{y}}(\pm i)J_{y}\eta_{i}^{z}\eta_{j}^{z}+\sum_{\langle ij\rangle_{z}}iJ_{z}\eta_{i}^{z}\eta_{j}^{z}.
\end{equation}
The Hamiltonian $\mathcal{H}^{''}$ has the same spectrum as the spectrum obtained in the Kitaev solution \cite{Kitaev2006} but for $\eta^{z}$ rather than for $c$ Majorana fermions (or in our definition the $\eta^{t}$ Majorana fermion). Note that in $\mathcal{H}^{''}$ the sign of the Majorana fermion $\eta^{z}$ is not a local degree freedom. Now, for the physical space to be the same we have to change the sign of two  Majorana fermions on the end-points of a given $z$-bond simultaneously. Mathematically, such transformation is given as $\eta_{i}^{\alpha}=\epsilon_{ij}\eta_{i}^{\alpha},\eta_{j}^{\alpha}=\epsilon_{ij}\eta_{j}^{\alpha}$ for every $z$-bond $\langle ij\rangle_z$ with $\epsilon_{ij}=\pm 1$. We call this a {\it reduced $Z_{2}$ gauge redundancy}.

\begin{figure}
\includegraphics[width=0.4\textwidth]{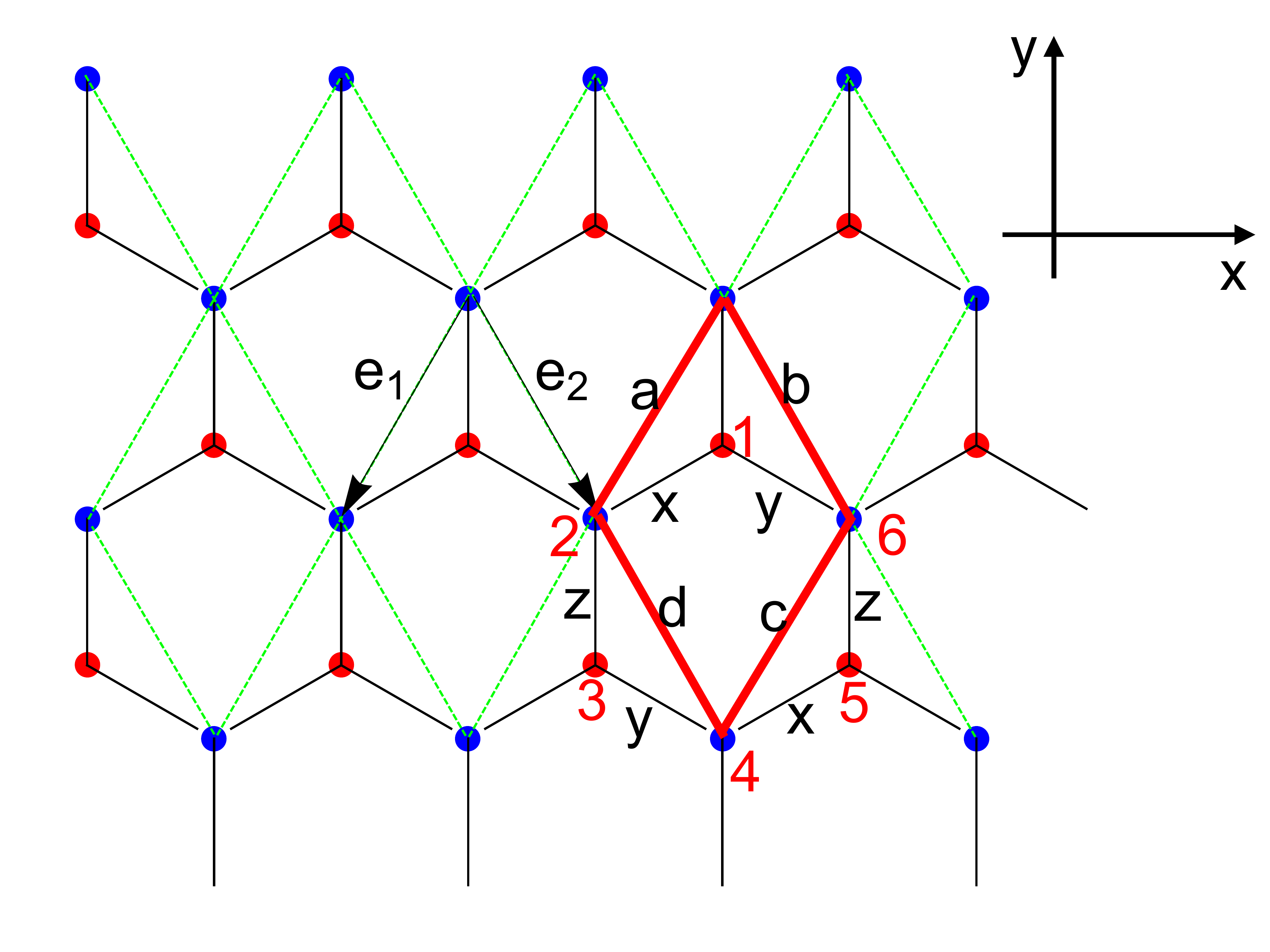}
\caption{The Kitaev model on the honeycomb lattice. $\boldsymbol{e}_{1}=a(-\frac{1}{2},-\frac{\sqrt{3}}{2}) $ and $\boldsymbol{e}_{2}=a(\frac{1}{2},-\frac{\sqrt{3}}{2}) $ are the primitive translations, with $a$ being the lattice constant. The two sublattices denoted in the main text by A and B are shown by blue and red dots, respectively. 
The sites of each of the two sublattices form a diamond lattice, as shown by the green dotted line for the A sublattice.}
\label{kitaevlattice}
\end{figure}

After the pairing of the two sites of every $z$-bond, we can define three complex fermions $c_{i}^{x},c_{i}^{y},c_{i}^{z}$ for each of the three flavors of Majorana fermions on every $z$-bond. \cite{Biswas2011} We use the honeycomb site of the A sublattice of the corresponding $z$-bond to label the real-space position of these complex fermions. Namely, we define
\begin{equation}
\label{definitionofcomplexc}
c_{i}^{\alpha}=\frac{1}{2}(\eta_{i}^{\alpha}+i\eta_{j}^{\alpha}), \qquad \alpha=x,y,z, 
\end{equation} 
for $z$-bond $\langle ij\rangle_{z}$ with $i$ in A sublattice and $j$ in B sublattice. Conversely, the Majorana fermion operators are given by $\eta_{i}^{\alpha}=c_{i}^{\alpha}+c_{i}^{\alpha\dagger}$ for site $i$ in the A sublattice, and $\eta_{j}^{\alpha}=-i(c_{i}^{\alpha}-c_{i}^{\alpha\dagger})$ for site $j$ in B sublattice. In the honeycomb lattice, the complex fermions live on the A sublattice and their position forms a {\it diamond lattice}, as shown by Fig.\ref{kitaevlattice}. From now on, we use vector $\boldsymbol{r}$ (and $\boldsymbol{r}'$) to label the sites of the diamond lattice while still using $i$ (and $j$) to label the sites of the honeycomb lattice.

By definition $\tau_{i}=-i\eta_{i}^{x}\eta_{i}^{y}\eta_{i}^{z}$, thus for every $z$-bond the condition (\ref{conditionofpair}) can be written as, \cite{Biswas2011}
\begin{equation}\label{constrcomplex}
\tau_{i}\tau_{j}\!=\!i(2n_{\boldsymbol{r}}^{x}\!-\!1)(2n_{\boldsymbol{r}}^{y}\!-\!1)(2n_{\boldsymbol{r}}^{z}\!-\!1)\!=\!-i(\!-1)^{n_{\boldsymbol{r}}^{x}+n_{\boldsymbol{r}}^{y}+n_{\boldsymbol{r}}^{z}}\! =\!-i,
\end{equation}
where $n_{\boldsymbol{r}}^{x}$, $n_{\boldsymbol{r}}^{y}$ and $n_{\boldsymbol{r}}^{z}$ are the bond fermion numbers on site $\boldsymbol{r}$ of the diamond lattice, corresponding to site $i$ of the honeycomb lattice. Thus, the requirement that for every $z$-bond $\tau_{i}\tau_{j}=-i$ is equivalent to the requirement that the total fermion number is  even on each  $z$-bond, or on each site of the diamond sublattice. Now for each diamond lattice site, all the states can be represented  as $|n^{x}_{\boldsymbol{r}},n^{y}_{\boldsymbol{r}},n^{z}_{\boldsymbol{r}}\rangle=|000\rangle, |110\rangle,|101\rangle,|011\rangle$. These four states span exactly the spin Hilbert space of the $z$-bond: $|\uparrow\uparrow\rangle$,$\frac{1}{\sqrt{2}}(|\uparrow\downarrow\rangle+|\downarrow\uparrow\rangle)$,$|\downarrow\downarrow\rangle$ and $\frac{1}{\sqrt{2}}(|\uparrow\downarrow\rangle-|\downarrow\uparrow\rangle)$. Thus, our Hilbert space corresponds to the physical Hilbert space. 
The other condition for solving the Hamiltonian can also be written in terms of the  $c$-fermions. Take, for example, the $x$-bond condition $i\eta_{i}^{y}\eta_{j}^{y}=\pm 1$, we then have
\begin{equation}
i\eta_{i}^{y}\eta_{j}^{y}=(c_{\boldsymbol{r}}^{y}c_{\boldsymbol{r}'}^{y}+c_{\boldsymbol{r}'}^{y\dagger}c_{\boldsymbol{r}}^{y\dagger})+(c_{\boldsymbol{r}}^{y\dagger}c_{\boldsymbol{r}'}^{y}+c_{\boldsymbol{r}'}^{y\dagger}c_{\boldsymbol{r}}^{y})=\pm 1,
\end{equation}
in which $\boldsymbol{r}$ and $\boldsymbol{r}'$ label the diamond lattice sites corresponding to sites $i$ and $j$ in the honeycomb lattice respectively.

With the complex fermions, one can obtain the energy spectrum of $\mathcal{H}^{''}$ in Eq. (\ref{hdoubleprime}), and show that it is the same as the one of the original Kitaev solution, \cite{Kitaev2006} see details in Appendix \ref{appendixcomplexfermion}.

\subsection{$Z_{2}$ Gauge Theory  and the Kitaev Model}

The simple solution in the previous section suffers from the following fact. In the solution, we take the energy eigenstate to be eigenstate of operator $\tau_{i}\tau_{j}$ on each $z$-bond, $\eta_{i}^{y}\eta_{j}^{y}$ on each x-bond and $\eta_{i}^{x}\eta_{j}^{x}$ on each $y$-bond, but the three groups of operators do not mutually commute. This means that the eigenstates found in this way are hardly the true eigenstates of the model itself. To remedy this problem, we start from the Hamiltonian (\ref{Ham1}) and try to map the Hamiltonian into a more familiar one. In doing so, we will find the proper eigenstates of the system and, thus, discuss how the solutions presented above are related to the real eigenstates. 

Firstly, let's define a bond operator
\begin{equation}
T_{ij}^{\alpha}\equiv i \eta_{i}^{\alpha}\eta_{j}^{\alpha},
\end{equation}
where $\alpha=x,y$. Note that once we transform into complex fermions, the lattice is diamond lattice with only $x$-bonds and $y$-bonds left, as shown in Fig \ref{kitaevlattice}. In terms of complex fermions, the bond operator is written as
\begin{equation}\label{Txyrr'}
T_{\boldsymbol{r}\boldsymbol{r}'}^{\alpha}=
c_{\boldsymbol{r}}^{\alpha}
c_{\boldsymbol{r}'}^{\alpha}+c_{\boldsymbol{r}'}^{\alpha\dagger}c_{\boldsymbol{r}}^{\alpha\dagger}+c_{\boldsymbol{r}}^{\alpha\dagger}c_{\boldsymbol{r}'}^{\alpha}+c_{\boldsymbol{r}'}^{\alpha\dagger}c_{\boldsymbol{r}}^{\alpha},
\end{equation}
where $\alpha=x,y$. To obtain the solution of the Kitaev model we have required that on x-bonds $T_{\boldsymbol{r}\boldsymbol{r}'}^{y}$ takes its eigenstates with eigenvalues $\pm 1$ and on y-bonds $T_{\boldsymbol{r}\boldsymbol{r}'}^{x}$ takes its eigenstates with eigenvalues $\pm 1$. 

Let us now give the complex fermions $c_{\boldsymbol{r}}^{x}$ or $c_{\boldsymbol{r}}^{y}$ a closer look. In the previous section we found that  $c_{\boldsymbol{r}}^{x}$ and $c_{\boldsymbol{r}}^{y}$ are completely independent. For a given diamond site, the number of these fermions can be $n_{\boldsymbol{r}}^{x}=0,1$ and $n_{\boldsymbol{r}}^{y}=0,1$ independently. For a given $y$-bond $\langle \boldsymbol{r}\boldsymbol{r}' \rangle$ on the diamond lattice, the Hilbert space can be defined by the occupation numbers $|n_{\boldsymbol{r}}^{x},n_{\boldsymbol{r}'}^{x}\rangle$, and there are four states $|00\rangle, |01\rangle, |10\rangle, |11\rangle$. The action of the operator $T_{\boldsymbol{r}\boldsymbol{r}'}^{x}$ gives the following: $T_{\boldsymbol{r}\boldsymbol{r}'}^{x}|00\rangle=|11\rangle, \quad T_{\boldsymbol{r}\boldsymbol{r}'}^{x}|01\rangle=|10\rangle,\quad T_{\boldsymbol{r}\boldsymbol{r}'}^{x}|10\rangle=|01\rangle, \text{ and } T_{\boldsymbol{r}\boldsymbol{r}'}^{x}|11\rangle=|00\rangle$. This is equivalent to say that the operator $T_{\boldsymbol{r}\boldsymbol{r}'}^{x}$ flips the occupation number of $c^{x}$ fermions on both sites $\boldsymbol{r}$ and $\boldsymbol{r}'$ independently. If we map the Hilbert space to a spin space and associate fermion occupation state $|1\rangle$ with spin state $|{\tilde \uparrow}\rangle$ and $|0\rangle$ with $|{\tilde \downarrow}\rangle$ on each site, the operator $T_{\boldsymbol{r}\boldsymbol{r}'}^{x}$ is equivalent to $\tilde{\tau}_{\boldsymbol{r}}^{x}\tilde{\tau}_{\boldsymbol{r}'}^{x}$, in which we use $\tilde{\tau}$ to label the new type of spin to avoid confusion with previous notations. Therefore we arrive at the following mapping
\begin{equation}
T_{\boldsymbol{r}\boldsymbol{r}'}^{\alpha}\rightarrow \{\tilde{\tau}_{\boldsymbol{r}}^{x}\tilde{\tau}_{\boldsymbol{r}'}^{x}\}^{\alpha},  \qquad \alpha=x,y,
\end{equation}
which is actually changing from one bosonic operator to another bosonic operator acting on the  Hilbert spaces of the same dimension. 

With this mapping the problem of the original Kitaev model has been changed to the following: on each site of the diamond lattice there is one complex fermion $c_{\boldsymbol{r}}^{z}$ interacting with two flavors of spins, $\tilde{\tau}_{1\boldsymbol{r}}^{\alpha}$ and $\tilde{\tau}_{2\boldsymbol{r}}^{\alpha}$: one flavor of spin interacts with the $c_{\boldsymbol{r}}^{z}$ fermions only on $x$-bonds and the other flavor interacts with $c_{\boldsymbol{r}}^{z}$ only on $y$-bonds. Having performed this reformulation, we now write the Hamiltonian $\mathcal{H}^{'}$ in Eq.(\ref{hprime}) in terms of these new variables as 
\begin{eqnarray}
\begin{aligned}
\label{hamiltoniannew}
\mathcal{H}^{'}=&\sum_{\boldsymbol{r}\in A}-J_{x}(\tilde{\tau}_{2\boldsymbol{r}}^{x}\tilde{\tau}_{2,\boldsymbol{r}+\boldsymbol{e}_{1}}^{x})[(c_{\boldsymbol{r}}^{z}+c_{\boldsymbol{r}}^{z\dagger})(c_{\boldsymbol{r}+\boldsymbol{e}_{1}}^{z}-c_{\boldsymbol{r}+\boldsymbol{e}_{1}}^{z\dagger})]\\
&-J_{y}(\tilde{\tau}_{1\boldsymbol{r}}^{x}\tilde{\tau}_{1,\boldsymbol{r}+\boldsymbol{e}_{2}}^{x})[(c_{\boldsymbol{r}}^{z}+c_{\boldsymbol{r}}^{z\dagger})(c_{\boldsymbol{r}+\boldsymbol{e}_{2}}^{z}-c_{\boldsymbol{r}+\boldsymbol{e}_{2}}^{z\dagger})]\\&+J_{z}(2c_{\boldsymbol{r}}^{z\dagger}c_{\boldsymbol{r}}^{z}-1).
\end{aligned}
\end{eqnarray}
Now  we use a duality transformation from the site spins to the bond spins for the two quasi-one-dimensional spin chains along $x$ and $y$-bonds of the diamond lattice:\cite{Kogut1979,fradkin2013field} 
\begin{equation}
\tilde{\tau}_{\boldsymbol{r}}^{x}\tilde{\tau}_{\boldsymbol{r}'}^{x}\rightarrow \tilde{\sigma}_{\boldsymbol{r}\boldsymbol{r}'}^{z},\, \tilde{\sigma}_{\boldsymbol{r}-\boldsymbol{e}_{1(2)},\boldsymbol{r}}^{x}\tilde{\sigma}_{\boldsymbol{r},\boldsymbol{r}+\boldsymbol{e}_{1(2)}}^{x}\rightarrow \tilde{\tau}_{\boldsymbol{r}}^{z},
\end{equation}
where $\boldsymbol{e}_{1}$ and $\boldsymbol{e}_{2}$ are used for $x$ and $y$-bonds of the diamond lattice, respectively. Since the new spin variables $\tilde{\sigma}_{\boldsymbol{r}\boldsymbol{r}'}^{z}$ are defined specifically on each type of the  bonds and thus are independent by nature, we can drop the indices 1 and 2 of the $\tilde{\tau}_{\boldsymbol{r}}^{x}$ operators. Now the Hamiltonian is written as follows
\begin{eqnarray}
\begin{aligned}
\label{hamiltonianz2}
\mathcal{H}^{'}=&\sum_{\boldsymbol{r}\in A}-J_{x}(\tilde{\sigma}_{\boldsymbol{r},\boldsymbol{r}+\boldsymbol{e}_{1}}^{z})[(c_{\boldsymbol{r}}^{z}+c_{\boldsymbol{r}}^{z\dagger})(c_{\boldsymbol{r}+\boldsymbol{e}_{1}}^{z}-c_{\boldsymbol{r}+\boldsymbol{e}_{1}}^{z\dagger})]\\
&-J_{y}(\tilde{\sigma}_{\boldsymbol{r},\boldsymbol{r}+\boldsymbol{e}_{2}}^{z})[(c_{\boldsymbol{r}}^{z}+c_{\boldsymbol{r}}^{z\dagger})(c_{\boldsymbol{r}+\boldsymbol{e}_{2}}^{z}-c_{\boldsymbol{r}+\boldsymbol{e}_{2}}^{z\dagger})]\\&+J_{z}(2c_{\boldsymbol{r}}^{z\dagger}c_{\boldsymbol{r}}^{z}-1).
\end{aligned}
\end{eqnarray}
This Hamiltonian describes a complex fermion on a diamond lattice interacting with a $Z_{2}$ gauge field defined on the bonds. Since the diamond lattice is equivalent to a square lattice,  known results for the  $Z_{2}$ gauge theory on the square lattice\cite{Kogut1979,fradkin2013field} can be borrowed here.

However, this mapping is not complete until we consider the constraints. Recall the original constraint that there are an even number of complex fermions on each diamond site (see Eq. (\ref{constrcomplex})). The fermion occupation numbers $n_{\boldsymbol{r}}^{x}$ and $n_{\boldsymbol{r}}^{y}$ in terms of two flavors of spin $\tilde{\tau}_{1,\boldsymbol{r}}$ and $\tilde{\tau}_{2,\boldsymbol{r}}$ are given by
\begin{equation}
n_{\boldsymbol{r}}^{x}=\frac{1}{2}(\tilde{\tau}_{1,\boldsymbol{r}}^{z}+1),\qquad n_{\boldsymbol{r}}^{y}=\frac{1}{2}(\tilde{\tau}_{2,\boldsymbol{r}}^{z}+1). 
\end{equation}
 Consequently, 
in terms of the new bond spins they can be written as:
\begin{equation}
\begin{aligned}
&n_{\boldsymbol{r}}^{x}=\frac{1}{2}(\tilde{\sigma}_{\boldsymbol{r}-\boldsymbol{e}_{1},\boldsymbol{r}}^{x}\tilde{\sigma}_{\boldsymbol{r},\boldsymbol{r}+\boldsymbol{e}_{1}}^{x}+1),\\&
n_{\boldsymbol{r}}^{y}=\frac{1}{2}(\tilde{\sigma}_{\boldsymbol{r}-\boldsymbol{e}_{2},\boldsymbol{r}}^{x}\tilde{\sigma}_{\boldsymbol{r},\boldsymbol{r}+\boldsymbol{e}_{2}}^{x}+1). 
\end{aligned}
\end{equation}
Considering that the fermion occupation numbers can only be 0 and 1, the constraint that there are even number of fermion per site can be written as 
\begin{eqnarray}
\label{constraints}
(-1)^{n_{\boldsymbol{r}}^{z}}\tilde{\sigma}_{\boldsymbol{r}-\boldsymbol{e}_{1},\boldsymbol{r}}^{x}\tilde{\sigma}_{\boldsymbol{r},\boldsymbol{r}+\boldsymbol{e}_{1}}^{x}\tilde{\sigma}_{\boldsymbol{r}-\boldsymbol{e}_{2},\boldsymbol{r}}^{x}\tilde{\sigma}_{\boldsymbol{r},\boldsymbol{r}+\boldsymbol{e}_{2}}^{x} &=&1, \\  \nonumber
(-1)^{n_{\boldsymbol{r}}^{z}}\prod_{\boldsymbol{r}'}\tilde{\sigma}_{\boldsymbol{r}\boldsymbol{r}'}^{x}& =&1, 
\end{eqnarray}
where $\boldsymbol{r}'$ are the four nearest neighbour sites to $\boldsymbol{r}$.

The model Hamiltonian (\ref{hamiltonianz2}) and the constraints (\ref{constraints}) define a model of complex matter fermions interacting with a $Z_{2}$ gauge field. The fermion number measures the defects of {\it star operators} in the system (See Fig.\ref{diamondlattice}). In terms of the lattice gauge theory the constraint (\ref{constraints}) is also interpreted as the {\it Gauss law}. \cite{fradkin2013field, Prosko2017} 

From now on, we will drop the index $z$ of the matter fermion whenever no confusion is caused. One important observation is that the operator in the constraints (\ref{constraints}) commutes with the Hamiltonian $\mathcal{H}^{'}$ in  Eq.(\ref{hamiltonianz2}), that is 
\begin{equation}
[(-1)^{n_{\boldsymbol{r}}}\tilde{\sigma}_{\boldsymbol{r}-\boldsymbol{e}_{1},\boldsymbol{r}}^{x}\tilde{\sigma}_{\boldsymbol{r},\boldsymbol{r}+\boldsymbol{e}_{1}}^{x}\tilde{\sigma}_{\boldsymbol{r}-\boldsymbol{e}_{2},\boldsymbol{r}}^{x}\tilde{\sigma}_{\boldsymbol{r},\boldsymbol{r}+\boldsymbol{e}_{2}}^{x},\mathcal{H}^{'}]=0.
\end{equation}
To prove this, we use the fact that $\{(-1)^{n_{\boldsymbol{r}}}, c_{\boldsymbol{r}}+c_{\boldsymbol{r}}^{\dagger}\}=0$, and $\{(-1)^{n_{\boldsymbol{r}}}, c_{\boldsymbol{r}}-c_{\boldsymbol{r}}^{\dagger}\}=0$, and for spin variables, $\{\sigma^{z},\sigma^{x}\}=0$. To proceed, we define the operator as
\begin{equation}
\mathcal{D}_{\boldsymbol{r}}=(-1)^{n_{\boldsymbol{r}}^{z}}\tilde{\sigma}_{\boldsymbol{r}-\boldsymbol{e}_{1},\boldsymbol{r}}^{x}\tilde{\sigma}_{\boldsymbol{r},\boldsymbol{r}+\boldsymbol{e}_{1}}^{x}\tilde{\sigma}_{\boldsymbol{r}-\boldsymbol{e}_{2},\boldsymbol{r}}^{x}\tilde{\sigma}_{\boldsymbol{r},\boldsymbol{r}+\boldsymbol{e}_{2}}^{x}.
\end{equation}
We have $[\mathcal{D}_{\boldsymbol{r}},\mathcal{H}^{'}]=0$ and $[\mathcal{D}_{\boldsymbol{r}},\mathcal{D}_{\boldsymbol{r}^{'}}]=0$, so the eigenstates of the Hamiltonian can also be eigenstates of the operator $\mathcal{D}_{\boldsymbol{r}}$. Actually, {\it $\mathcal{D}_{\boldsymbol{r}}$ generates the $Z_{2}$ local gauge transformation of the model at site $\boldsymbol{r}$}, so the condition (\ref{constraints}) is equivalent to the gauge invariance of the states, \cite{fradkin2013field}
\begin{equation}
\label{gausscondition}
\mathcal{D}_{\boldsymbol{r}}|\psi\rangle_{\text{phys}}=|\psi\rangle_{\text{phys}}.
\end{equation}

\subsection{Physical States of the Model}

To describe the states of the pure $Z_{2}$ gauge theory without matter fermion, it is convenient to work in the $\sigma^{x}$ basis \cite{fradkin2013field}, resulting in a geometric intepretation of the states in terms of loops. With matter fermion as in our model, it is more useful to work in the $\sigma^{z}$ basis, which will have a close relationship with the Kitaev solution. First, we consider the transformed Hamiltonian (\ref{hamiltonianz2}). Without considering the Gauss-law condition (\ref{gausscondition}), the {\it naive} eigenstates can be written as
\begin{equation}
\label{naiveeigenstates}
|\psi\rangle=|\{\sigma_{\boldsymbol{r}\boldsymbol{r}^{'}}^{z}\}\rangle\otimes|\phi_{\{\sigma^{z}\}}\rangle, \qquad \mathcal{H}|\psi\rangle=E|\psi\rangle.
\end{equation}
In this state $|\{\sigma_{\boldsymbol{r}\boldsymbol{r}^{'}}^{z}\}\rangle$ denotes the product state of eigenstate of $\sigma_{\boldsymbol{r}\boldsymbol{r}^{'}}^{z}$ on each bond, for every such distribution, $\mathcal{H}$ will reduce to a free fermion Hamiltonian for complex fermion $c$, with eigenstate $|\phi_{\{\sigma^{z}\}}\rangle$ corresponding to the distribution $\{\sigma^{z}\}$. The Kitaev solution, although given in a different approach, is simply one of the states $|\psi\rangle$. 

Now, we enforce the Gauss-law condition (\ref{gausscondition}). We note again that the $\mathcal{D}_{\boldsymbol{r}}$ operators commute with each other and with the Hamiltonian and generates the local gauge transformation. Since $\mathcal{D}_{\boldsymbol{r}}^{2}=1$, we have $(\mathcal{D}_{\boldsymbol{r}}-1)\frac{1+\mathcal{D}_{\boldsymbol{r}}}{2}|\psi\rangle=0$ and thus we can define the following projection of states from the eigenstates (\ref{naiveeigenstates}):
\begin{equation}
\label{projectedstate}
\hat{\mathcal{P}}|\psi\rangle=2^{\frac{N-1}{2}}(\prod_{\boldsymbol{r}}\frac{1+\mathcal{D}_{\boldsymbol{r}}}{2})|\psi\rangle=\frac{1}{2^{\frac{N+1}{2}}}(\sum_{\{\boldsymbol{r}\}}\prod_{\boldsymbol{r}^{'}\in\{\boldsymbol{r}\}}\mathcal{D}_{\boldsymbol{r}^{'}})|\psi\rangle.
\end{equation}
This projected state is given by equal superposition of all the states in the same gauge sector as $|\psi\rangle$ as shown in the last equation of (\ref{projectedstate}), and the prefactor $2^{\frac{N-1}{2}}$ is added to ensure the proper normalization of the states. This projected state satisfies two properties. First, since $\mathcal{D}_{\boldsymbol{r}}$ commutes with each other, we have $\mathcal{D}_{\boldsymbol{r}}\hat{\mathcal{P}}|\psi\rangle=\hat{\mathcal{P}}|\psi\rangle$, that is, the Gauss-law condition (\ref{gausscondition}) is satisfied. Second, since $\mathcal{D}_{\boldsymbol{r}}$ commutes with $\mathcal{H}$, we have $\mathcal{H}\hat{\mathcal{P}}|\psi\rangle=\hat{\mathcal{P}}\mathcal{H}|\psi\rangle=E\hat{\mathcal{P}}|\psi\rangle$, which means that the projected state (\ref{projectedstate}) is eigenstate of the Hamiltonian with the same energy as the unprojected one (\ref{naiveeigenstates}). 

If the system has periodic boundary conditions, i.e. defined on a torus, then the operation of performing gauge transformation for all the sites is important. For the operator $\hat{\mathcal{P}}$ to be non-zero, this operation has to have eigenvalue +1 instead of -1 for every physical state. This means that 
\begin{equation}
\prod_{\boldsymbol{r}}(-1)^{n_{\boldsymbol{r}}}\prod_{\boldsymbol{r}^{'}}\tilde{\sigma}^{x}_{\boldsymbol{r}\boldsymbol{r}^{'}}=(-1)^{\sum_{\boldsymbol{r}}n_{\boldsymbol{r}}}=+1,
\end{equation}
that is, the total number of fermion has to be an even number. 

In order to calculate physical observables, we note that the projected states are the {\it physical states} of the model. However, the unprojected state $|\psi\rangle$ given by (\ref{naiveeigenstates}) can sometimes be useful as well, in terms of energy spectra, they give the same result. For other gauge invariant operators $\hat{\mathcal{O}}$, we have $[\hat{\mathcal{O}},\hat{\mathcal{P}}]=0$, thus
\begin{equation}
\label{physicalobservables}
\langle \hat{\mathcal{O}}\rangle=\langle \psi|\hat{\mathcal{P}}\hat{\mathcal{O}}\hat{\mathcal{P}}|\psi\rangle=\langle \psi|\hat{\mathcal{O}}\hat{\mathcal{P}}^{2}|\psi\rangle=2^{\frac{N-1}{2}}\langle \psi|\hat{\mathcal{O}}\hat{\mathcal{P}}|\psi\rangle,
\end{equation}
in which we use the fact that $\hat{\mathcal{P}}^{2}=2^{\frac{N-1}{2}}\hat{\mathcal{P}}$.


\subsection{Generalized $Z_2$ gauge theory with complex matter fermions}

Finally we want to generalize the model, in particular Eq.(\ref{hamiltonianz2}), by adding the standard terms discussed in the context of $Z_2$ lattice gauge theories, \cite{Prosko2017,Kogut1979,fradkin2013field} such that
\begin{eqnarray}
\label{HamGeneral}
\mathcal{H}=&\mathcal{H}^{'} + \mathcal{H}_g.
\end{eqnarray}


\begin{figure}
\centering
\includegraphics[width=70mm]{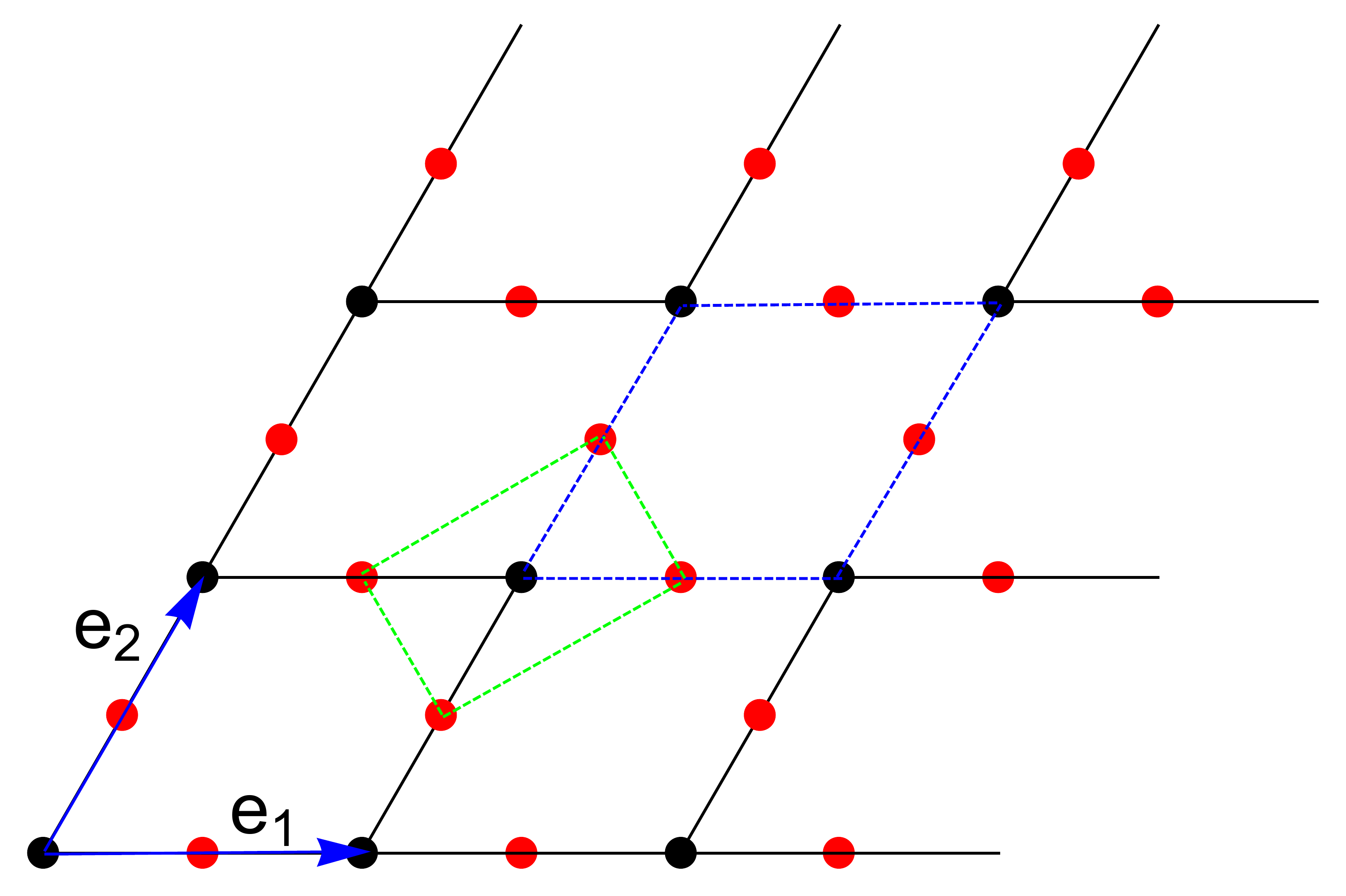}
\caption{The diamond lattice corresponding to the original honecomb lattice, with unit vectors $\boldsymbol{e}_{1},\boldsymbol{e}_{2}$. The matter (complex) fermion lives on the lattice sites, labelled by the black dots. The $Z_{2}$ gauge connection $\tilde{\sigma}^{z}$ lives on the bonds, labelled by the red dots. One of the {\it plaquettes} is shown by the blue dotted lines and one of the {\it star operators} used in the constraint (\ref{constraints}) is shown by the green dotted lines.}
\label{diamondlattice}
\end{figure}

Let us discuss the individual terms of the pure gauge part of the Hamiltonian, $\mathcal{H}_{g}=\mathcal{H}_{P}+\mathcal{H}_{E}$, separately. The first term is given by the the magnetic plaquette term 
\begin{equation}
\label{plaquetteterm}
\mathcal{H}_{P}=-K\sum_{P}\prod_{\boldsymbol{r}\boldsymbol{r}^{'}\in\partial P}\tilde{\sigma}_{\boldsymbol{r}\boldsymbol{r}^{'}}^{z},
\end{equation}
which turns out to be related to the plaquette operator $W_{p}=\sigma_{1}^{x}\sigma_{2}^{y}\sigma_{3}^{z}\sigma_{4}^{x}\sigma_{5}^{y}\sigma_{6}^{z}$ defined in the original Kitaev model \cite{Kitaev2006}, see Fig. \ref{kitaevlattice}. To see this correspondence, we note that the bond spin $\tilde{\sigma}^{z}_{\boldsymbol{r}\boldsymbol{r}'}$ defined on the diamond lattice actually comes from the product of two Majorana fermion on the corresponding honecomb bond, in particular $\tilde{\sigma}_{\boldsymbol{r},\boldsymbol{r}+\boldsymbol{e}_{1}}^{z}=-i\eta_{\boldsymbol{r}+\boldsymbol{e}_{1},A}^{y}\eta_{\boldsymbol{r},B}^{y}$, for x-bond; and $\tilde{\sigma}_{\boldsymbol{r},\boldsymbol{r}+\boldsymbol{e}_{2}}^{z}=-i\eta_{\boldsymbol{r}+\boldsymbol{e}_{2},A}^{x}\eta_{\boldsymbol{r},B}^{x}$ for y-bond. (In these expressions, we use $\boldsymbol{r}$ to label the sites of diamond lattice and $\boldsymbol{r}+\boldsymbol{e}_{1},A$ and $\boldsymbol{r},B$ denote the two sites of the x-bond of the honeycomb lattice belonging to A and B sublattice). Using this, we have for one plaquette operator (the labelling of the sites is shown in Fig. \ref{kitaevlattice} and the plaquette is shown explicitly in Fig.  \ref{diamondlattice}),
\begin{eqnarray}
\begin{aligned}
\tilde{\sigma}_{a}^{z}\tilde{\sigma}_{b}^{z}\tilde{\sigma}_{c}^{z}\tilde{\sigma}_{d}^{z}&=(-i\eta_{2}^{y}\eta_{1}^{y})(-i\eta_{6}^{x}\eta_{1}^{x})(-i\eta_{4}^{y}\eta_{5}^{y})(-i\eta_{4}^{x}\eta_{3}^{x})\\&=\sigma_{1}^{z}\sigma_{2}^{y}\sigma_{3}^{x}\sigma_{4}^{z}\sigma_{5}^{y}\sigma_{6}^{x}=W_{P}.
\end{aligned}
\end{eqnarray}
In this equation, we have used the condition that for every z-bond $ij$, $\tau_{i}\tau_{j}=-i$ and the representation for spin operators (\ref{so3majoranarepresentation}) and (\ref{so3majoranarep2}). Thus, the addition of the plaquette term (\ref{plaquetteterm}) is just adding a term of $W_{P}$ in the original Hamiltonian,
\begin{equation}
-K\sum_{P}W_{P}\leftrightarrow \mathcal{H}_{P}.
\end{equation}
We note that this magnetic plaquette term has been considered in a generalized Kitaev model proposed in Ref. \onlinecite{Schaffer2012}, in which mean field theory is applied to study the quantum phases of the generalized Kitaev model.

The second term of $\mathcal{H}_{g}$ should be the electric part $\mathcal{H}_{E}=-h\sum_{\boldsymbol{r}\boldsymbol{r}^{'}}\tilde{\sigma}_{\boldsymbol{r}\boldsymbol{r}^{'}}^{x}$, which can be shown to correspond to the discrete lattice version of the standared Electromagnetic Hamiltonian \cite{Prosko2017} $\mathcal{H}\propto E^{2}$. However, this term renders the whole system non-integrable preventing an exact solution. Therefore, we follow Prosko {\it et al.} \cite{Prosko2017} and add an alternative electric Hamiltonian, which corresponds to a standared Electromagnetic Hamiltonian $\mathcal{H} \propto (\partial_{\mathbf{r}}E)^{2}$. It is given by the Kitaev star operator in the Toric code model \cite{Kitaev2003} 
\begin{eqnarray}
\mathcal{H}_{E}=-h\sum_{\boldsymbol{r}}\prod_{\boldsymbol{r}^{'}}\tilde{\sigma}_{\boldsymbol{r}\boldsymbol{r}^{'}}^{x}, 
\end{eqnarray}
with $\boldsymbol{r}^{'}$ being the four sites adjecent to $\boldsymbol{r}$. Crucially, this term can be translated to a shift of the  $J_{z}$ term in the Hamitonian (\ref{hamiltonianz2}) potential term due to the Gauss-law constraint (\ref{constraints})
\begin{eqnarray}
\begin{aligned}
&J_{z}(2c_{\boldsymbol{r}}^{\dagger}c_{\boldsymbol{r}}\!-\!1)\!-\!h\sum_{\boldsymbol{r}}\prod_{\boldsymbol{r}^{'}}\tilde{\sigma}_{\boldsymbol{r}\boldsymbol{r}^{'}}^{x} 
\leftrightarrow (J_{z}\!+\!h)(2c_{\boldsymbol{r}}^{\dagger}c_{\boldsymbol{r}}\!-\!1).
\end{aligned}
\end{eqnarray}

Overall, our exactly soluble generalized Hamiltonian for a $Z_{2}$ gauge field interacting with matter fermions takes the form
\begin{eqnarray}
\label{fullhamiltonian}
\begin{aligned}
\mathcal{H}=&\mathcal{H}^{'} -K\sum_{P}\prod_{\boldsymbol{r}\boldsymbol{r}^{'}\in\partial P}\tilde{\sigma}_{\boldsymbol{r}\boldsymbol{r}^{'}}^{z}-h\sum_{\boldsymbol{r}}\prod_{\boldsymbol{r}^{'}}\tilde{\sigma}_{\boldsymbol{r}\boldsymbol{r}^{'}}^{x}.
\end{aligned}
\end{eqnarray}
In the equation above, the P denotes the plaquettes in the two-dimensional lattice. The full Hamiltonian (\ref{fullhamiltonian}) together with the Guass law condition (\ref{gausscondition}) defines a general $Z_{2}$ gauge field theory interacting with complex matter fermion. It can be solved exactly taking the simplified form
\begin{widetext}
\begin{eqnarray}
\label{fullhamiltonian1}
\begin{aligned}
\mathcal{H}
=&\!\sum_{\boldsymbol{r}\in A}\!-\!J_{x}(\tilde{\sigma}_{\boldsymbol{r},\boldsymbol{r}\!+\!\boldsymbol{e}_{1}}^{z})[(c_{\boldsymbol{r}}\!+\!c_{\boldsymbol{r}}^{\dagger})(c_{\boldsymbol{r}\!+\!\boldsymbol{e}_{1}}\!-\!c_{\boldsymbol{r}\!+\!\boldsymbol{e}_{1}}^{\dagger})]
\!-\!J_{y}(\tilde{\sigma}_{\boldsymbol{r},\boldsymbol{r}\!+\!\boldsymbol{e}_{2}}^{z})[(c_{\boldsymbol{r}}\!+\!c_{\boldsymbol{r}}^{\dagger})(c_{\boldsymbol{r}\!+\!\boldsymbol{e}_{2}}\!-\!c_{\boldsymbol{r}\!+\!\boldsymbol{e}_{2}}^{\dagger})]\!+\!(J_{z}\!+\!h)(2c_{\boldsymbol{r}}^{\dagger}c_{\boldsymbol{r}}\!-\!1)\!-\!K\!\sum_{P}\!W_P.
\end{aligned}
\end{eqnarray}
\end{widetext}

\begin{figure}
\centering
\includegraphics[width=70mm]{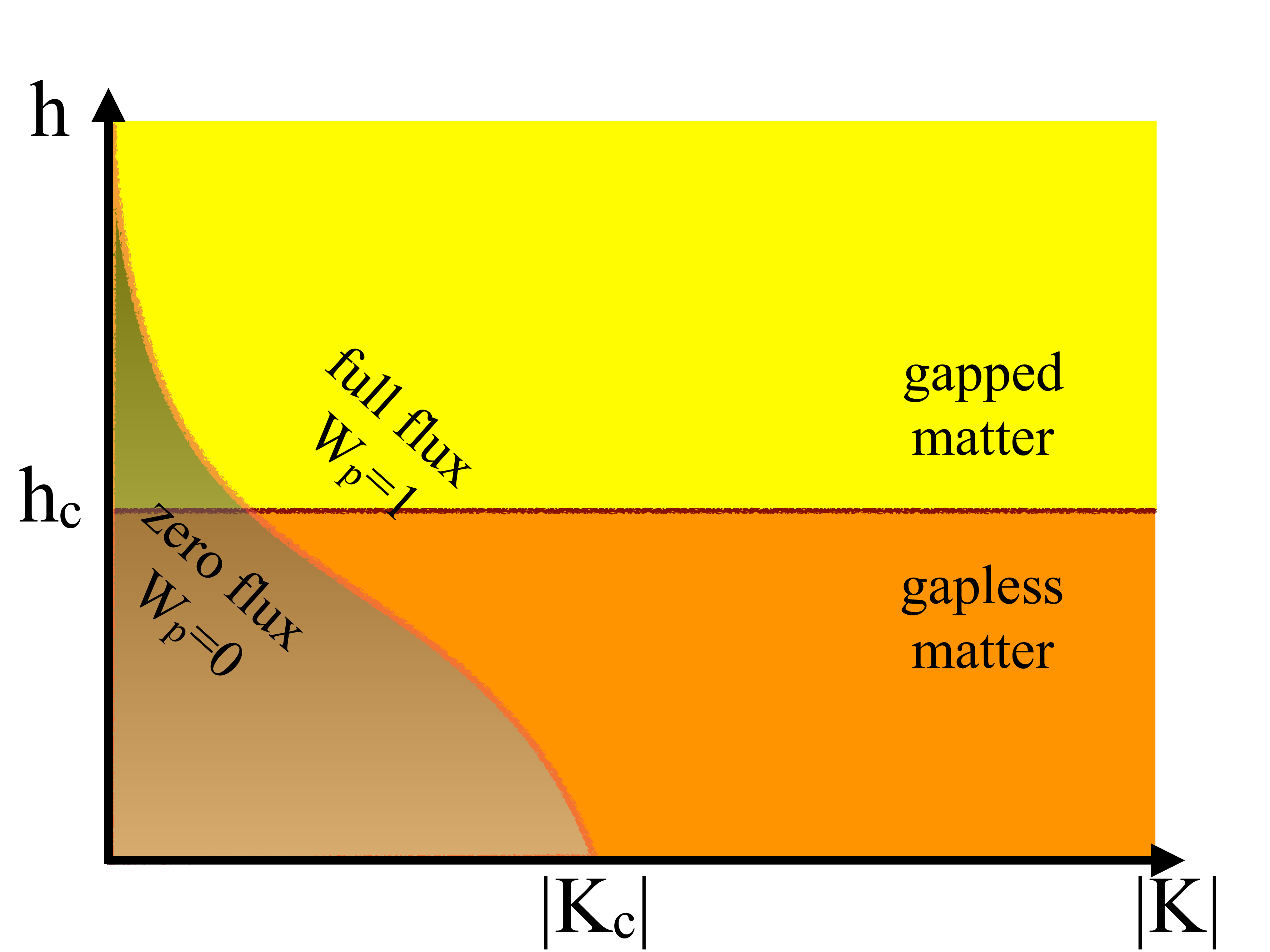}
\caption{The schematic ground state phase diagram of the generalized $Z_2$ gauge theory with matter fermions given by Eq. (\ref{fullhamiltonian}) is shown for isotropic couplings $J_x=J_y=J_z$ as a function of $K<0$ and $h$.}
\label{phasediagram}
\end{figure}

From the exact solution of the Kitaev model we can directly read off the influence of the two different gauge terms. Without the magnetic plaquette term $K=0$ the ground state is flux free with $W_P=+1$ for all plaquettes $P$. In the presence of finite coupling, there will be a critical $K_c<0$ below which the full flux state state with $W_P=-1$ for all $P$ has lower energy. The electric term leads to an anisotropy of the coupling constants such that for $|(J_z+h)|>|J_x|+|J_y|$ the fermionic spectrum is gapped both for the zero\cite{Kitaev2006} and full flux sector.\cite{Lahtinen2012} Note, an increase in the effective anisotropy also reduced the gap between flux sectors. Therefore, we sketch the overall ground state phase diagram of the Hamiltonian Eq. (\ref{fullhamiltonian}) for isotropic exchange ($J_x=J_y=J_z$) in Fig. \ref{phasediagram}. Note, the transition between the full and zero flux sector might be via intermediate flux configurations.

\section{Conclusion and Outlook} \label{secconclusion}

We have presented the correspondence between the three types of Majorana representation of spin and the spinor representation of SO(4) which helps to also see the connection to the well known complex fermion representation (\ref{complexslaverep}). The fact that Majorana fermions satisfies the same Clifford algebra as the Gamma matrices in the Dirac spinor representation leads us to the decomposation of the spin operators into bilinears of Majorana fermions following a similar decomposation as of SO(4) generators into Gamma matrices. To achieve the spin SU(2) algebra, the SO(3) Majorana representation corresponds to the SO(3) subgroup of SO(4). On the other hand, the Kitaev and the SO(4) chiral Majorana representation use the $\text{SU(2)}_{L}$ subgroup of SO(4). The complex fermion representation is equivalent to the SO(4) chiral Majorana representation and there is a mapping between its Hilbert spaces. \cite{Mohapatra80,Wilczek82,Chetan1996} Looking ahead, it is desirable to explore more rigorously the general mapping between SO($2n$) Dirac spinor spaces and the Hilbert space of $n$ complex fermions \cite{Mohapatra80,Wilczek82,Chetan1996,Ahlbrecht2009} in connection with slave-particle descriptions of quantum spin systems.

It is important to emphasize that the Majorana Hilbert space of the SO(3) representation is different from the other two in that it requires pairing of sites to give the correct correspondence to the spin space. However, it has the big advantage that no unphysical states are involved, thus any spin Hamiltonian is faithfully represented by the Majorana fermions. This enables new solutions to spin models which have already been studied by other representations. In this paper, we used the SO(3) Majorana representation to obtain an alternative solution of the celebrated Kitaev model, which maps it to a $Z_{2}$ gauge theory interacting with matter fermions. Our choice of the Kitaev model is unique because previous results on the SO(3) Majorana representation, such as Refs. \onlinecite{Herfurth2013,Shastry1997,Biswas2011}, were concentrated on the mean-field study of QSL states. It is an interesting direction for future research to consider other types of exactly solvable models using this spin representation or to investigate whether its structure permits the construction of new soluble models.

Our solution of the Kitaev model in Sec \ref{seckitaevmodel} highlights the role of the $Z_{2}$ gauge transformation in relating the physical states and the naive eigenstates of the Hamiltonian. We give an explicit formula for the calculation of physical observables using the naive eigenstates, namely Eq. (\ref{physicalobservables}), which is also applicable for small system sizes away from the thermodynamic limit.\cite{Pedrocchi2011,zschocke2015} Using this, one can calculate experimental observables, e.g. the spin structure factor, or the finite temperature behavior.\cite{Nasu2014} These calculations are left for future work.

Finally, we have shown that our solution of the Kitaev model takes the form of a special $Z_2$ lattice gauge theory coupled to matter fermions, Eq. (\ref{hamiltonianz2}). We have generalized the latter by adding the usual gauge field terms\cite{fradkin2013field,Kogut1979,senthil2000}, e.g. a magnetic plaquette term, and an electric star term\cite{Prosko2017} to the Hamiltonian (\ref{fullhamiltonian}). Our generalized lattice gauge theory with matter fermions can be solved exactly via the Gauss law constraint and we have sketched its phase diagram as a function of the new coupling constant. It is an interesting direction for future research to explore the phase diagram beyond the integrable limit and to study its quantum phase transitions.

\section*{Acknowledgements}

JF thanks M. Voloshin for insightful discussions. JK acknowledges helpful conversations with A. Smith.  NP and JF acknowledge the support from NSF DMR-1511768 Grant.

\appendix

\section{Mapping between the Majorana Hilbert space and the Dirac spinor space} \label{appendixmapping}

From the similarity between the Majorana fermions and the $\gamma$ matrices discussed in Sec. \ref{secmajoranaspinor}, there is a one-to-one mapping from the Hilbert space of $f_{\uparrow}$ and $f_{\downarrow}$ to the Dirac spinor space, the basis of the latter is decomposed into left-handed and right-handed Weyl spinor $\psi_{D}=(\psi_{L}^{T},\psi_{R}^{T})$, with the basis $\psi_{L}:(1,0)^{T},(0,1)^{T}$ and $\psi_{R}:(1,0)^{T},(0,1)^{T}$ . The natural choice of this mapping would be (define $|\downarrow\uparrow\rangle=f_{\downarrow}^{\dagger}f_{\uparrow}^{\dagger}|0\rangle$)
\begin{eqnarray}
\begin{aligned}
&|\uparrow\rangle\sim\left(
\begin{array}{c}
\left(
\begin{array}{c}
1\\0
\end{array}
\right)\\\left(
\begin{array}{c}
0\\0
\end{array}
\right)
\end{array}
\right); 
|\downarrow\rangle\sim\left(
\begin{array}{c}
\left(
\begin{array}{c}
0\\1
\end{array}
\right)\\\left(
\begin{array}{c}
0\\0
\end{array}
\right)
\end{array}
\right); \\  
&|\downarrow\uparrow\rangle\sim\left(
\begin{array}{c}
\left(
\begin{array}{c}
0\\0
\end{array}
\right)\\\left(
\begin{array}{c}
1\\0
\end{array}
\right)
\end{array}
\right);
|0\rangle\sim\left(
\begin{array}{c}
\left(
\begin{array}{c}
0\\0
\end{array}
\right)\\\left(
\begin{array}{c}
0\\1
\end{array}
\right)
\end{array}
\right).
\end{aligned}
\end{eqnarray}
Under such mapping the projection to left-handed spinors is equivalent to the projection to single-occupation fermion states. With this mapping at hand, we see that the linear operation of the fermion Hilbert space is mapped onto the linear operation of the Dirac spinor space. In particular, the four Majorana fermions correspond to four matrices in the spinor space, 
\begin{eqnarray}
\label{majoranamatrices}
\begin{aligned}
&\eta^{t}\sim \left(
\begin{array}{cc}
0&-i\sigma^{x}\\i\sigma^{x}&0
\end{array}
\right) \quad \eta^{x}\sim \left(
\begin{array}{cc}
0&I_{2}\\I_{2}&0
\end{array}
\right)\\ &\eta^{y}\sim \left(
\begin{array}{cc}
0&-i\sigma^{z}\\i\sigma^{z}&0
\end{array}
\right) \quad \eta^{z}\sim \left(
\begin{array}{cc}
0&i\sigma^{y}\\-i\sigma^{y}&0
\end{array}
\right).
\end{aligned}
\end{eqnarray}

It is well-known that the complex fermion representation of spin has a SU(2) gauge redundancy, \cite{Lee2006} which is given by
\begin{equation}
\label{complexgauge}
\left(\begin{array}{c}
f_{\uparrow}\\f_{\downarrow}^{\dagger}
\end{array}\right)\rightarrow \mathcal{U}\left(\begin{array}{c}
f_{\uparrow}\\f_{\downarrow}^{\dagger}
\end{array}\right), \qquad \mathcal{U}\in \text{SU(2)}.
\end{equation}
This operation leaves the spin representation (\ref{complexslaverep}) invariant. Due to the one-to-one correspondence between the complex fermion representation and the SO(4) chiral Majorana fermion representation (\ref{so4chiralrep}), there should be a SU(2) gauge redundancy in the latter as well, this is given by
\begin{equation}
\label{chiralmajoranagauge}
\boldsymbol{\eta}\rightarrow \mathcal{U}\boldsymbol{\eta},\,{\text {\rm where}\,\,}   \mathcal{U}\,{\text {\rm are}} \,4\times 4\,{\text {\rm \,real\, matrices}}.
\end{equation}

A general SU(2) gauge transformation in Eq. (\ref{complexgauge}) can be written as $\,\mathcal{U}=a_{0}I+i\boldsymbol{a}\cdot\boldsymbol{\sigma}$, where $a_{0}$ and $\boldsymbol{a}$ are real and satisfy $a_{0}^{2}+|\boldsymbol{a}|^{2}=1$, and $\boldsymbol{\sigma}$ are the Pauli matrices. With this, we can express the matrix $\mathcal{U}$ in (\ref{chiralmajoranagauge}) as
\begin{eqnarray}
\mathcal{U}=\left(\begin{array}{cccc}
a_{0}&-a_{1}&a_{2}&-a_{3}\\a_{1}&a_{0}&-a_{3}&-a_{2}\\-a_{2}&a_{3}&a_{0}&-a_{1}\\a_{3}&a_{2}&a_{1}&a_{0}
\end{array}\right)
\end{eqnarray}
With this matrix $\mathcal{U}$, the transformation (\ref{chiralmajoranagauge}) leaves the SO(4) chiral Majorana representation invariant. 

Having defined  the mapping between the complex fermion Hilbert space and the SO(4) Dirac spinor, it is a natural question to ask if there is any correspondence between the SU(2) gauge invariance in the complex fermion representation and some symmetry in the Dirac spinor space. This question is hard to answer because the representation itself is not linear in the complex fermions. Explorations in this direction are left for future study.

\section{Complex fermion spectrum of Kitaev model} \label{appendixcomplexfermion}

In terms of complex fermions introduced in Sec. \ref{secsolutionofkitaev}, namely, Eq. (\ref{definitionofcomplexc}), the Hamiltonian (\ref{hdoubleprime}) becomes
\begin{eqnarray}
\begin{aligned}
\label{hdoubleprime2}
\mathcal{H}^{''}=&\sum_{\boldsymbol{r}\in A}(-\tilde{J}_{x})[(c_{\boldsymbol{r}}^{z}+c_{\boldsymbol{r}}^{z\dagger})(c_{\boldsymbol{r}+\boldsymbol{e}_{1}}^{z}-c_{\boldsymbol{r}+\boldsymbol{e}_{1}}^{z\dagger})]\\&+(-\tilde{J}_{y})[(c_{\boldsymbol{r}}^{z}+c_{\boldsymbol{r}}^{z\dagger})(c_{\boldsymbol{r}+\boldsymbol{e}_{2}}^{z}-c_{\boldsymbol{r}+\boldsymbol{e}_{2}}^{z\dagger})]\\&+J_{z}(2c_{\boldsymbol{r}}^{z\dagger}c_{\boldsymbol{r}}^{z}-1),
\end{aligned}
\end{eqnarray}
in which $\tilde{J}_{x}=\pm J_{x},\tilde{J}_{y}=\pm J_{y}$ depending on the eigenvalues of $\eta_{i}^{y}\eta_{j}^{y}$ and $\eta_{i}^{x}\eta_{j}^{x}$  on $x$- and  $y$-bonds, respectively. In particular, the $+$ signs correspond to the flux free ground state sector with all $W_p=+1$.

Next we perform a Fourier transformation of $c$ fermions 
\begin{equation}\label{FT}
c_{\boldsymbol{r}}=\frac{1}{\sqrt{N}}\sum_{\boldsymbol{k}}c_{\boldsymbol{k}}e^{i\boldsymbol{k}\cdot\boldsymbol{r}}; \qquad c_{\boldsymbol{r}}^{\dagger}=\frac{1}{\sqrt{N}}\sum_{\boldsymbol{k}}c_{\boldsymbol{k}}^{\dagger}e^{-i\boldsymbol{k}\cdot\boldsymbol{r}},
\end{equation}
where N denotes the total number of unit cells in the system. This sum in the k-space is taken over the entire first Brillouin Zone, labelled by BZ, for the Hamiltonian only half of the k-points are independent and we want to sum over half of the BZ, labelled by $BZ^{'}$, so the Hamiltonian (\ref{hdoubleprime2}) is transformed into:
\begin{eqnarray}\label{Hmomentum}
\begin{aligned}
\mathcal{H}^{''}=\sum_{\boldsymbol{k}\in BZ^{'}}
\left(\begin{array}{cc}
c_{-\boldsymbol{k}}^{z}&c_{\boldsymbol{k}}^{z\dagger}
\end{array}\right)
\mathcal{M}_{\boldsymbol{k}}\left(\begin{array}{c}\begin{aligned}
& c_{-\boldsymbol{k}}^{z\dagger}\\& c_{\boldsymbol{k}}^{z}
\end{aligned}
\end{array}\right),
\end{aligned}
\end{eqnarray}
in which the coupling matrix $\mathcal{M}_{\boldsymbol{k}}$ is given by
\begin{widetext}
\begin{equation}
\mathcal{M}_{\boldsymbol{k}}=\left(\begin{array}{cc}
\begin{aligned}
&-2\tilde{J}_{x}\cos(\boldsymbol{k}\cdot\boldsymbol{e}_{1})-2\tilde{J}_{y}\cos(\boldsymbol{k}\cdot\boldsymbol{e}_{2})-2J_{z}& 2i(\tilde{J}_{x}\sin(\boldsymbol{k}\cdot\boldsymbol{e}_{1})+\tilde{J}_{y}\sin(\boldsymbol{k}\cdot\boldsymbol{e}_{2}))\\&-2i(\tilde{J}_{x}\sin(\boldsymbol{k}\cdot\boldsymbol{e}_{1})+\tilde{J}_{y}\sin(\boldsymbol{k}\cdot\boldsymbol{e}_{2}))& 2\tilde{J}_{x}\cos(\boldsymbol{k}\cdot\boldsymbol{e}_{1})+2\tilde{J}_{y}\cos(\boldsymbol{k}\cdot\boldsymbol{e}_{2})+2J_{z}
\end{aligned}
\end{array}\right).
\end{equation} 
\end{widetext} 
   
Diagonalize this matrix, we obtain the energy spectrum for the $c^{z}$ fermions which agrees with the one obtained in the Kitaev's solution,\cite{Kitaev2006}
\begin{equation}
E_{\boldsymbol{k}}=2|\tilde{J}_{x}e^{i\boldsymbol{k}\cdot\boldsymbol{e}_{1}}+\tilde{J}_{y}e^{i\boldsymbol{k}\cdot\boldsymbol{e}_{2}}+J_{z}|.
\end{equation}


\bibliography{refJFUnew}
\end{document}